\documentclass[fleqn,usenatbib]{mnras}

\usepackage{newtxtext,newtxmath}

\usepackage[T1]{fontenc}

\DeclareRobustCommand{\VAN}[3]{#2}
\let\VANthebibliography\thebibliography
\def\thebibliography{\DeclareRobustCommand{\VAN}[3]{##3}\VANthebibliography}

\usepackage{graphicx}	
\usepackage{amsmath}	
\usepackage{booktabs}
\usepackage{multirow}
\usepackage{makecell}
\usepackage{array}
\usepackage{xcolor}
\usepackage{gensymb}
\usepackage{multirow}
\usepackage{amsmath}
\usepackage{enumitem}
\usepackage{placeins}
\usepackage{soul}
\usepackage{float}

\hypersetup{
	colorlinks=true,
	linktoc=all,
	linkcolor=blue,
}
\newcommand{\orcid}[1]{\href{#1}{\includegraphics[scale=0.04]{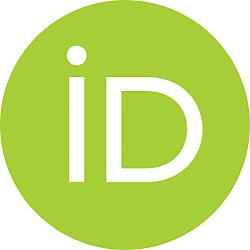}}}

\newcommand\dens{\text{cm}^{-3}}
\newcommand\fZero{\texttt{f00}}
\newcommand\fLow{\texttt{f01}}
\newcommand\fHigh{\texttt{f10}}
\newcommand\fZeroB{\texttt{f00\_M120}}
\newcommand\fZeroC{\texttt{f00\_M370}}
\newcommand\msun{\text{M}_\odot}

\title[Impact of a Clumpy ISM on AGN Jets]{The Impact of a Clumpy Ambient Medium on the Dynamics \\ and Synchrotron Emission of AGN Jets}

\author[Almeida, Fendt \& Vaidya]{
Ivan Almeida,$^{1}$\thanks{E-mail: ivalmeida@mpia.de}\orcid{https://orcid.org/0000-0001-6018-2852}
Christian Fendt,$^{1}$\thanks{E-mail: fendt@mpia.de}\orcid{https://orcid.org/0000-0002-3528-7625}
Bhargav Vaidya$^{2}$\thanks{E-mail: bvaidya@iiti.ac.in}\orcid{https://orcid.org/0000-0001-5424-0059}
\\
$^{1}$Max Planck Institute for Astronomy, Königstuhl 17, D-69117 Heidelberg, Germany\\
$^{2}$ Department of Astronomy, Astrophysics and Space Engineering, Indian Institute of Technology Indore, Khandwa Road, Simrol Indore 453552, India\\
}

\date{Accepted 2026 September 23. Received 2026 September 17; in original form 2026 July 07.}

\pubyear{\the\year{}}

\begin{document}
\label{firstpage}
\pagerange{\pageref{firstpage}--\pageref{lastpage}}
\maketitle

\begin{abstract}
Relativistic jets from AGN are expected to propagate through an inhomogeneous ISM, but the role of small-scale gas inhomogeneities in shaping their early evolution remains uncertain. We investigate this problem using 3D special-relativistic magnetohydrodynamic simulations with \texttt{PLUTO}, following a jet with a Lorentz factor of 10 propagating through a pc-scale medium. We compare a homogeneous ambient medium with two clumpy ambient-medium models, motivated by dense gas in the nuclear ISM, in which clouds occupy volume filling factors of $0.1$ and $1$~per cent.
We compute synthetic synchrotron maps and spectra from Lagrangian macro-particles that have been accelerated by diffusive shock acceleration.
Jet--cloud interactions deflect the flow, enhance shock formation, and produce a more asymmetric cocoon than in the homogeneous ISM case. Altogether, this results in a substantially stronger mass loading: the entrained mass increases from $0.43\,\msun$ in the homogeneous run to $5.3$ and $8.8\,\msun$ in the clumpy runs. The clumpy ISM also modifies the turbulence and the material mixing within the jet cocoon.
The dynamical differences we find translate into brighter and more irregular synchrotron emission.
For the same injected jet, the models with a clumpy medium produce stronger frequency-integrated synchrotron emission than the homogeneous run, with the largest differences in the SED occurring from the sub-mm to the infrared/optical bands.
Our results demonstrate that including a small-scale ISM structure provides a more complete description of the early dynamical evolution of young AGN jets and their radiative properties.
\end{abstract}

\begin{keywords}
galaxies: jets -- MHD -- relativistic processes
\end{keywords}

\section{Introduction}\label{sec:intro}

Relativistic jets are observed in a wide range of astrophysical systems, including active galactic nuclei (AGN), X-ray binaries, and gamma-ray bursts \citep{Blandford2019}. 
They are launched in the vicinity of accreting compact objects, where magnetic fields play a central role in launching and collimating relativistic outflows \citep{Blandford1977, dalPino2005, Hawley2016}. 

AGN jets are among the most powerful manifestations of black hole activity, and can propagate from the vicinity of the central supermassive black hole (SMBH) to kiloparsec or megaparsec scales \citep{Marscher2011}. Their non-thermal emission extends across the electromagnetic spectrum, from radio \citep{Lister2016, Saikia2022} to $\gamma$-ray energies \citep{Madejski2016, deMenezes2020}. Regardless of their observed manifestation, AGN jets originate from accreting SMBHs and propagate through the gas surrounding the galactic nucleus, where their interaction with the interstellar medium (ISM) can affect their dynamics, morphology, and emission.

A central uncertainty in modelling AGN jet propagation is the structure of the ambient gas through which the jet advances. While numerical simulations commonly adopt smooth or homogeneous backgrounds, the gas in galactic nuclei is expected to be highly inhomogeneous, with dense structures embedded within a more diffuse component \citep{Davies2020, Revalski2025}. 
Simulations of AGN jets propagating through clumpy or multiphase environments have shown that dense structures can obstruct and redirect the jet, alter its morphology and advance speed, enhance coupling to the surrounding gas, and modify the evolution of the inflated bubble or cocoon \citep{Sutherland2007, Wagner2012, Meenakshi2022, Mukherjee2018}. 
Complementary simulations of AGN-driven winds and outflows have further demonstrated that ambient inhomogeneities affect gas acceleration and inferred mass and momentum outflow rates \citep{Wagner2013, Ward2024, Almeida2026}. 
However, the effect of controlled variations in the cloud volume filling factor on the early, pc-scale evolution of a relativistic magnetised jet remains insufficiently quantified, particularly when its propagation, entrainment and mixing, cocoon turbulence, and spatially resolved synchrotron emission are considered together.

As AGN jets propagate through their surroundings, shocks, shear layers, and turbulent structures can develop within the beam and cocoon, affecting both the flow dynamics and its non-thermal emission. 
Shocks provide sites where particles may be accelerated to relativistic energies through diffusive shock acceleration \citep{Hargrave1974, Upreti2024, Elley2026, Costa2026}, while additional energisation may arise from stochastic acceleration, magnetic reconnection, or velocity shear \citep{Kundu2021, Kadowaki2021, Sironi2021, Das2025}. 
Numerical studies have therefore increasingly combined relativistic jet dynamics with non-thermal particle evolution to connect internal flow structures and magnetic-field variations with spatially resolved emission signatures \citep{Mukherjee2021, Borse2021, Dubey2023, Dubey2024}.
In a clumpy ambient medium, jet--cloud interactions may modify the locations and strengths of shocks, distort the magnetic-field structure, and redistribute the emitting jet material. 
Synthetic synchrotron emission therefore provides an observable diagnostic of the dynamical imprint of ambient inhomogeneities on relativistic jets.

In this work, we perform three-dimensional special-relativistic magnetohydrodynamic (RMHD) simulations with the \texttt{PLUTO} code \citep{Mignone2007, Mignone2011} to investigate the propagation of a relativistic magnetised jet through an idealised clumpy ISM on parsec scales. 
We compare a homogeneous ambient medium with clumpy media characterised by different cloud volume filling factors, motivated by observational constraints on the density and volume filling factor of dense gas in the narrow-line region of nearby AGN.
To model the associated non-thermal emission, we inject Lagrangian macro-particles into the jet using the particle module developed by \citet{Vaidya2018}, following the setup of \citet{Dubey2023, Dubey2024} (hereafter \textit{D23} and \textit{D24}). Each macro-particle represents an ensemble of non-thermal electrons that is advected with the flow and evolves through shock acceleration, adiabatic losses, synchrotron cooling, and inverse-Compton losses due to cosmic microwave background photons. 
We compute the synchrotron emissivity of these particles and integrate it along the line of sight to produce synthetic emission maps. 
In addition, passive scalar tracers allow us to distinguish jet- and cloud-origin material and to quantify the effects of ambient clumpiness on mixing, entrainment, mass loading, and the dynamical evolution of the jet and cocoon.

The paper is organised as follows. In Section~\ref{sec:model}, we describe the numerical setup, including the jet injection, the clumpy ambient-medium model, the non-thermal particle treatment, and the calculation of synchrotron emission. 
In Section~\ref{sec:jet-dynamics}, we examine the dynamical evolution of the jet and quantify the effect of ambient clumpiness on its structure and mass entrainment. 
In Section~\ref{sec:material-mixing}, we analyse jet--ISM mixing and momentum transfer. 
Section~\ref{sec:synchrotron} presents the resulting synchrotron signatures and discusses their observational interpretation. 
In Sections~\ref{sec:leptons-baryons} and~\ref{sec:turbulence}, we discuss the implications for the distribution of jet-associated leptons and entrained baryonic material, and for turbulence in the cocoon, respectively. 
We summarise the limitations of the present model and future extensions in Section~\ref{sec:limitations}, before presenting our conclusions in Section~\ref{sec:summary}.
\section{Model Setup}\label{sec:model}

\subsection{Numerical Framework}\label{subsec:code}
In order to investigate the jet dynamics and evolution, we apply
the \texttt{PLUTO} code \citep{Mignone2007, Mignone2011} for solving the related 3D-RMHD fluid equations,
\begin{equation}
\frac{\partial}{\partial t} \left( \begin{array}{c}
D \\
\mathbf{m} \\
\mathbf{B} \\
E_t 
\end{array} \right) + \nabla \cdot \left( \begin{array}{c}
D \mathbf{v} \\
w_t \gamma_f^2 \mathbf{vv} - \mathbf{bb} + \overline{\textbf{I}}P_t  \\
\mathbf{vB} - \mathbf{Bv} \\
\mathbf{m} 
\end{array} \right)^{T} = 0.
\label{eq:RMHD-equations}
\end{equation}
Here, $D = \rho\gamma_f$, $\rho$ is the proper rest-mass density, $\gamma_f$ the Lorentz factor of the fluid, $\mathbf{v}$ the velocity, $\mathbf{m}$ the momentum, $w_t = \rho h + b_m^2$ the relativistic total enthalpy, $\mathbf{B}$ the magnetic field in the laboratory frame, $b_m^2 = \mathbf{B}^2/\gamma_f^2 + (\mathbf{v} \cdot \mathbf{B})^2$, $\mathbf{b} = \mathbf{B}/\gamma + \gamma (\mathbf{v} \cdot \mathbf{B})\mathbf{v}$, $E_t$ the total energy density, $P_t = P + b^2_m/2$ the total pressure, and $\overline{\mathbf{I}}$ the identity tensor. 

For the equation of state, we adopt the Taub–Mathews (TM) approximation \citep{Mathews1971, Mignone2005}, where the specific enthalpy $h$ is given by
\begin{equation}
    h = \frac{5}{2} \Theta + \sqrt{\frac{9}{4} \Theta^2 + 1},
\label{eq:Taub–Mathews-EoS}
\end{equation}
with $\Theta=p/\rho$ the dimensionless temperature. For further details on the equations, see Section 2 of \textit{D23}.

The computational domain is a Cartesian box extending from $-5$ to $5$~pc in both transverse directions and from $0$ to $20$~pc along the jet-propagation axis\footnote{The simulations are performed in code units with $c=1$, $n_{\rm bg} = 1 \, {\rm cm}^{-3}$, and a length unit $L_0=0.5$~pc, corresponding to the jet radius. The dimensional scaling can therefore be adjusted by rescaling $n_{\rm bg}$ and $L_0$, provided the dimensionless parameters are preserved.}.
We use an equidistant Cartesian grid (x,y,z) of $501 \times 501 \times 1001$ grid cells.
We apply outflow boundary conditions at all domain boundaries, imposing zero gradients on the primitive variables.
The jet-injection procedure follows \textit{D23} and is summarised in Section~\ref{subsec:jet}. 
The simulations are evolved for $t\simeq85$~yr, stopping before the jet head reaches close to the upper boundary at $z=20$~pc.

\subsection{Jet Injection} \label{subsec:jet}
A relativistic jet is continuously injected throughout the simulation through a cylindrical nozzle placed inside the computational domain. The nozzle occupies the cylindrical region $r=\sqrt{x^2+y^2}\leq r_j$, with
$r_j=0.5$~pc, and extends over $0\leq z \leq 0.5$~pc.
Within this region, the primitive variables are reset at each timestep to impose a steady inflow.

The injected beam follows the relativistic, rotating, magnetised equilibrium solution of \citet{Bodo2019}. 
This prescription sets the transverse profiles of the velocity and magnetic-field components.
The injected configuration is transversely force-free, preventing an initial artificial expansion associated with an over-pressurised beam.
For convenience, the full set of equations used in the nozzle is given in Appendix~\ref{app:jet-profile}.

The jet is injected with density\footnote{Throughout this paper, number densities refer to hydrogen number densities, assuming a hydrogen mass fraction of 76 per cent.} 
$n_j=0.01$~cm$^{-3}$, pressure $P_j=1.49\times10^{-6}$~dyne cm$^{-3}$, central axial magnetic field $B_{zc}=2.4$~mG, central pitch angle $\delta_c=0.01$, angular velocity $\Omega_c = 0.4 \, \times (c / r_j) $, Lorentz factor $\gamma_c=10$, and magnetisation radius $a=0.3$~pc, and total jet power of $4.3 \times 10^{44}$~erg s$^{-1}$. The nozzle-averaged plasma-$\beta$ of the injected beam is $0.25$. These parameters are the same as those adopted in the steady nozzle setup of \textit{D23} (see their Table~1).

\subsection{ISM model} \label{subsec:ISM-model}
We model the ambient medium as an idealised representation of a clumpy nuclear ISM. 
It consists of a diffuse background gas and a population of embedded dense clouds. 
This prescription is not intended to capture the full multiphase structure, chemistry,  or turbulence of the ISM, but isolates the effect of dense gas inhomogeneities on jet propagation.
We define the ISM as all material initially located outside the jet-injection nozzle. The ISM is modelled as a two-component medium, consisting of a diffuse background gas and a population of embedded dense clouds.

All ISM material is initially at rest, $\mathbf{v}_{\rm ISM}=0$. 
We choose the initial cloud thermal pressure to match the background gas thermal pressure, $P_{\rm cl}=P_{\rm bg}$, to avoid artificial early cloud expansion or collapse.
The initial ISM magnetic field is purely axial, with $B_z^{\rm ISM}=2.4$~mG and $B_x^{\rm ISM}=B_y^{\rm ISM}=0$.
The diffuse component has density $n_{\rm bg}=1\,\dens$ and pressure $p_{\rm bg}=1.49\times10^{-6}$~dyne cm$^{-3}$.
Initially, this gas fills the computational domain except for the injection nozzle and the regions occupied by dense clouds.

In the simulations with the clumpy ISM, the dense phase is represented by spherical clouds. The cloud radii span $R_{\rm cl}=0.05$--$1$~pc and are sampled from a power-law distribution, $N(R_{\rm cl})\propto R_{\rm cl}^{-2}$. The central density of each cloud, $n_c$, is drawn from a lognormal distribution, while the internal density profile decreases smoothly from $n_c$ at the cloud centre to the background density, $n_{\rm bg}$, at the cloud boundary. Further details of the cloud construction are given in Appendix~\ref{app:clouds}.

Observational estimates based on emission lines such as H$\alpha$ and [O~{\sc ii}] suggest dense-gas filling factors below $1$ per cent, with typical values around $0.1$ per cent \citep{Storchi2007, Collins2009}. Since these estimates trace relatively dense gas, we apply the density threshold $n>10^2\,\dens$, when defining filling factor $f$.
For each clumpy ISM simulation, the cloud population is normalised such that the volume occupied by gas denser than $10^2\,\dens$ matches the target dense-gas filling factor. We define
\begin{equation}
    f = \frac{V(n > 10^2 \, \dens)}{V_{\rm tot}}.
\label{eq:filling-factor}
\end{equation}

\begin{figure}
\includegraphics[width=\columnwidth]{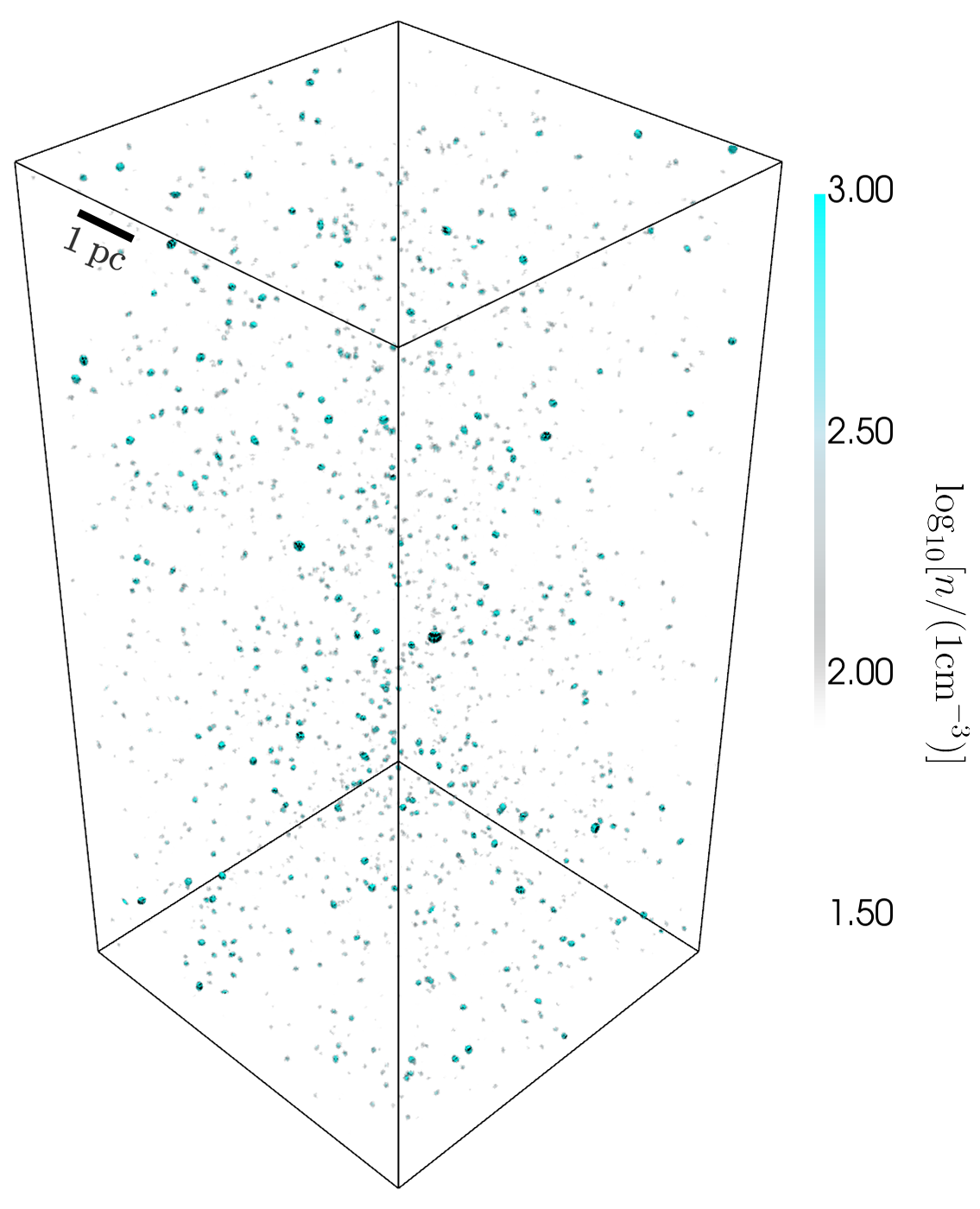}
\caption{Initial gas distribution for the \fLow\ simulation. The domain is filled with diffuse background gas with $n=1\,\dens$, within which denser clouds are embedded. 
The highlighted regions indicate gas with $n>10^2\,\dens$, which defines the dense-gas filling factor. 
The jet is injected from the lower boundary within a radius of $0.5$~pc.}
\label{fig:IC}
\end{figure}

Figure~\ref{fig:IC} shows the initial ISM structure for the \fLow\ simulation.
The diffuse background fills most of the domain, while the highlighted regions show gas with $n>10^2\,\dens$, which defines the dense-gas filling factor.

We perform three main simulations: a homogeneous reference run with $f=0$ (\fZero), a clumpy ISM run with $f=0.1$ per cent (\fLow), and a denser clumpy ISM run with $f=1$ per cent (\fHigh). 
We also run two homogeneous comparison simulations, \fZeroB\ and \fZeroC, with $f=0$ but with increased uniform
background densities chosen to match the total ISM masses of \fLow\ and \fHigh, respectively. 
These runs allow us to separate the effect of total ambient mass from the effect of spatial inhomogeneity. 
The simulation parameters are summarised in Table~\ref{tab:sims}.

\begin{table}
\centering
\begin{tabular}{c|cc}
Sim ID & $f$ & $M_{\rm ISM}$ [$M_\odot$] \\ \hline \hline
\fZero & $0.0\%$ & $60$  \\
\fLow & $0.1\%$  & $120$ \\
\fHigh & $1.0\%$ & $370$ \\ \hline \hline 
\fZeroB & $0.0\%$ & $120$\\
\fZeroC & $0.0\%$ & $370$\\ \hline \hline
\end{tabular}
\caption{Summary of the simulation suite. The table lists the dense-gas filling
factor $f$, defined by equation~\eqref{eq:filling-factor}, and the total initial
ISM mass within the computational domain. The runs \fZeroB\ and \fZeroC\ are
homogeneous comparison simulations with increased background densities chosen to
match the total ISM masses of \fLow\ and \fHigh, respectively.}
\label{tab:sims}
\end{table}

\subsection{Particle Injection} \label{subsec:particle-injection}

We use the Lagrangian particle module of \texttt{PLUTO} \citep{Vaidya2018, Mukherjee2021} to inject macroparticles from the jet nozzle, following \textit{D23} and \textit{D24}. 
The particles are advected with the local fluid velocity and are assigned the fluid Lorentz factor $\gamma_f$ of their host cell. They are treated as passive tracers and do not exert feedback on the RMHD evolution.

Each macroparticle represents an ensemble of non-thermal electrons whose initial particle distribution of micro-particles (i.e., leptons or electrons in your case) within each macro-particle is $N(\gamma) = \mathcal{N}_0 \gamma^{-\alpha}$, where $\gamma$ is the electron Lorentz factor --distinct from the fluid Lorentz factor $\gamma_f$-- and $\mathcal{N}_0$ is the normalisation.
We adopt $\alpha=6$ and restrict the distribution to $\gamma_{\rm min}=10^2$ and $\gamma_{\rm max}=10^8$.

\begin{equation}
    N_e = \int_{\gamma_{\rm min}}^{\gamma_{\rm max}}
    \mathcal{N}_0 \gamma^{-\alpha}\,d\gamma .
\end{equation}

The electron density can be calculated considering equipartition of the energy densities between the magnetic field and the radiating electrons 

\begin{equation}
N_{e} = \frac{\epsilon^{2}}{m_{e} c^{2}} \, \frac{B_{\mathrm{dyn}}^{2}}{2} 
\left( \frac{2 - \alpha}{1 - \alpha} \right) 
\left( \frac{\gamma_{\max}^{1 - \alpha} - \gamma_{\min}^{1 - \alpha}}
{\gamma_{\max}^{2 - \alpha} - \gamma_{\min}^{2 - \alpha}} \right).
\label{eq:Ne-equipartition}
\end{equation}
Here, $\epsilon=B_{\rm eq}/B_{\rm dyn}$, where $B_{\rm eq}$ is the magnetic field corresponding to equipartition and $B_{\rm dyn}$ is the nozzle-averaged total magnetic-field strength computed from the injected magnetic-field profile (we refer to \textit{D23} for a more detailed explanation), $m_e$ is the electron mass, $c$ is the speed of light. For our simulations, we assume $\epsilon = 0.01$, which is equivalent to $N_e \approx 0.003 \ \dens$.

The subsequent evolution of the electron distribution and the calculation of the synchrotron emissivity are described in Section~\ref{subsec:synchrotron}.

\subsection{Synchrotron-emission model}
\label{subsec:synchrotron}

The injected macroparticles are evolved following the prescription of \textit{D23} and \textit{D24}. Their electron energy distributions are modified by adiabatic losses, synchrotron cooling, IC--CMB losses, and shock acceleration through diffusive shock acceleration. 
For the compact pc-scale systems studied here at redshift $z=0$, IC--CMB losses are subdominant over the simulated time-scales, but they are retained for consistency with the particle-evolution model.

At each output time, we compute the synchrotron emissivity from the evolved electron distribution and the local magnetic field. 
We focus on synchrotron emission and do not include IC--CMB emission, synchrotron self-Compton emission, or external Compton emission from photon fields associated with the broad-line region, dusty torus, stellar radiation, or the extragalactic background. 
The detailed emissivity calculation follows \textit{D23} and \textit{D24} and is summarised in Appendix~\ref{app:synchrotron}.

\section{Dynamical evolution}\label{sec:jet-dynamics}

\subsection{Jet Propagation and Morphology}\label{subsec:jet-structure}

\begin{table}
\centering
\begin{tabular}{l|cccccc}
Sim ID & $f$ & $z_{\rm j}$ & $v_{\rm head}/c$ & $|\mathbf{B}|_{\rm med}$  & $|\mathbf{B}|_{90}$ \\
& & [pc] & & [mG] & [mG] \\ \hline \hline
\fZero\  & $0.0\%$ & 16 & 0.61 & 8.7 & 13.6 \\
\fLow\ & $0.1\%$ & 14 & 0.53 & 7.2 & 11.5 \\
\fHigh & $1.0\%$ & 11 & 0.42 & 9.8 &  14.7 \\ \hline \hline
\fZeroB & $0.0\%$ & 14 & 0.53 & 10.7 & 17.8 \\
\fZeroC &$0.0\%$ & 9 & 0.35 & 12.5 & 21.7 \\ \hline \hline
\end{tabular}
\caption{
Jet properties at $t_{\rm sim}=85$~yr.
The jet length, $z_{\rm j}$, is measured from the injection region to the jet head, and the average head velocity is defined as $v_{\rm head}\equiv z_{\rm j}/t_{\rm sim}$.
The last two columns give the median and 90th-percentile values of the magnetic-field magnitude.
}
\label{tab:sims-jet}
\end{table}

\begin{figure}
\includegraphics[width=0.95\columnwidth]{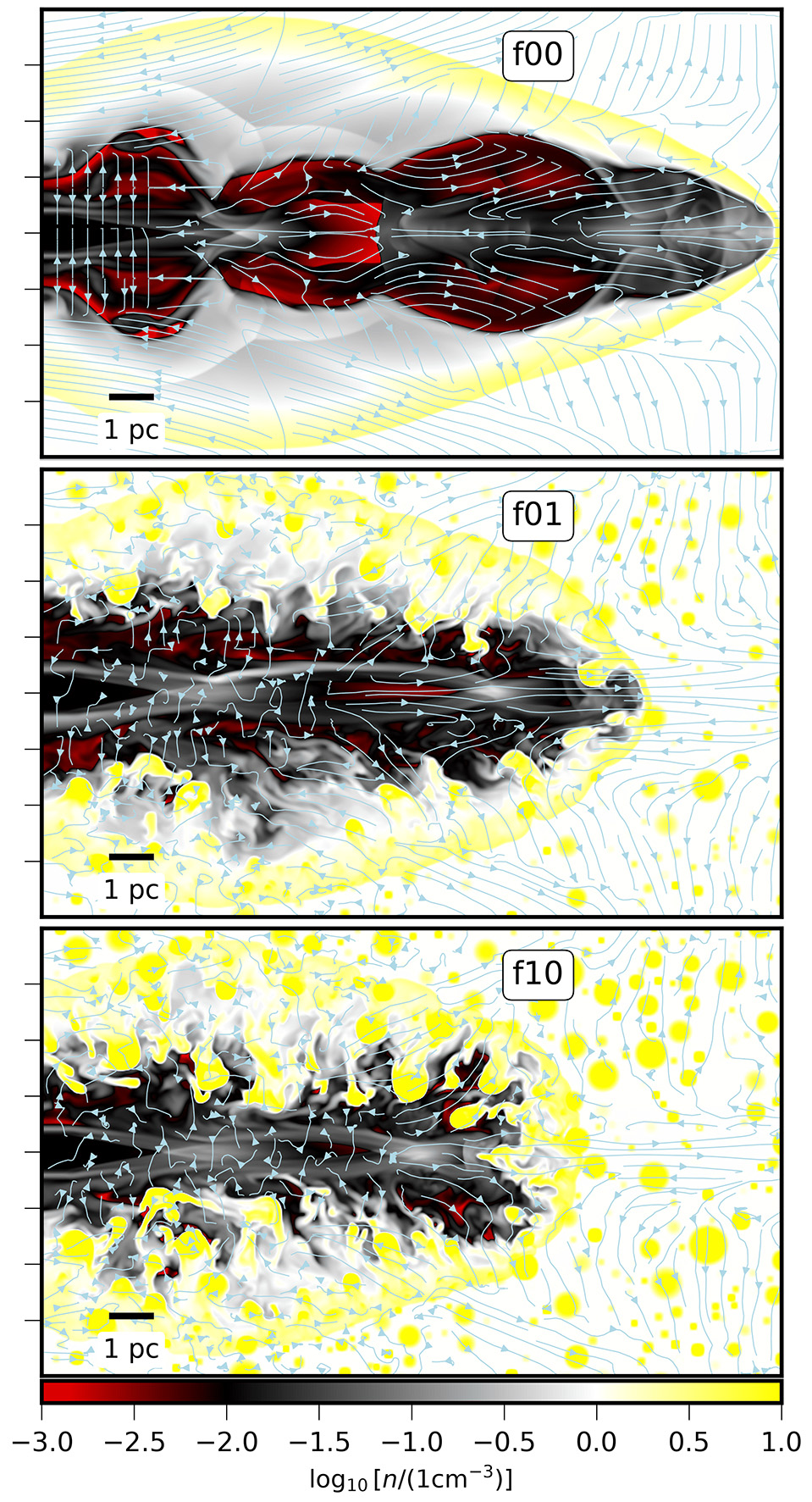}
\caption{
Logarithmic density maps overlaid with velocity streamlines at $t_{\rm sim}=85$~yr.
Panels correspond to \fZero\ (top), \fLow\ (middle), and \fHigh\ (bottom).
The homogeneous case shows a narrow, symmetric jet and a well-defined bow shock.
In the simulations with a clumpy medium, dense clouds locally deflect and distort the jet, producing a broader and more irregular cocoon.
Low-density channels between clouds allow part of the flow to propagate through the inhomogeneous medium.
}
\label{fig:jet-rho}
\end{figure}

Figure~\ref{fig:jet-rho} shows logarithmic density maps overlaid with velocity streamlines at $t_{\rm sim}=85$~yr for the three main simulations. These maps illustrate how the ambient density structure affects the propagation and morphology of the relativistic jet.

In the homogeneous case, \fZero, the jet maintains a highly symmetric morphology. The low-density jet material propagates efficiently along the $z$-axis, forming a narrow channel surrounded by a cocoon. The jet head drives a well-defined bow shock into the ambient medium, while secondary shocks develop within the beam as the flow adjusts to the surrounding gas.

The simulations considering a clumpy medium, \fLow\ and \fHigh, show a markedly different morphology. Dense clouds provide localised obstacles to the jet propagation, deflecting the flow and reducing the coherence of the jet structure. As a result, the cocoon becomes broader and more irregular, and the interface between the jet and the ambient medium is less sharply defined. This effect increases with the cloud filling factor, with \fHigh\ showing the strongest distortion and the most asymmetric cocoon morphology.

The denser ambient media slow the advance of the jet head. At $t_{\rm sim}=85$~yr, the jet reaches $z_{\rm j}\approx16$~pc in \fZero, compared to $z_{\rm j}\approx14$~pc in \fLow\ and $z_{\rm j}\approx11$~pc in \fHigh. The homogeneous comparison runs help separate the effect of total gas mass from that of the density structure. In the low-mass case, \fLow\ and \fZeroB\ have similar jet lengths and head velocities, indicating that the mean gas content largely controls the average jet advance. In the high-mass case, however, \fHigh\ propagates farther than \fZeroC, with $v_{\rm head}=0.42c$ compared to $0.35c$ (Table~\ref{tab:sims-jet}). This difference arises because, in the clumpy medium, part of the jet can propagate through low-density channels between clouds, whereas in the homogeneous comparison run there are no comparable low-density channels. Thus, the total gas mass affects the mean jet propagation speed, but the spatial distribution of this gas determines how efficiently the jet can channel through the ambient medium.

\begin{figure}
\includegraphics[width=\linewidth]{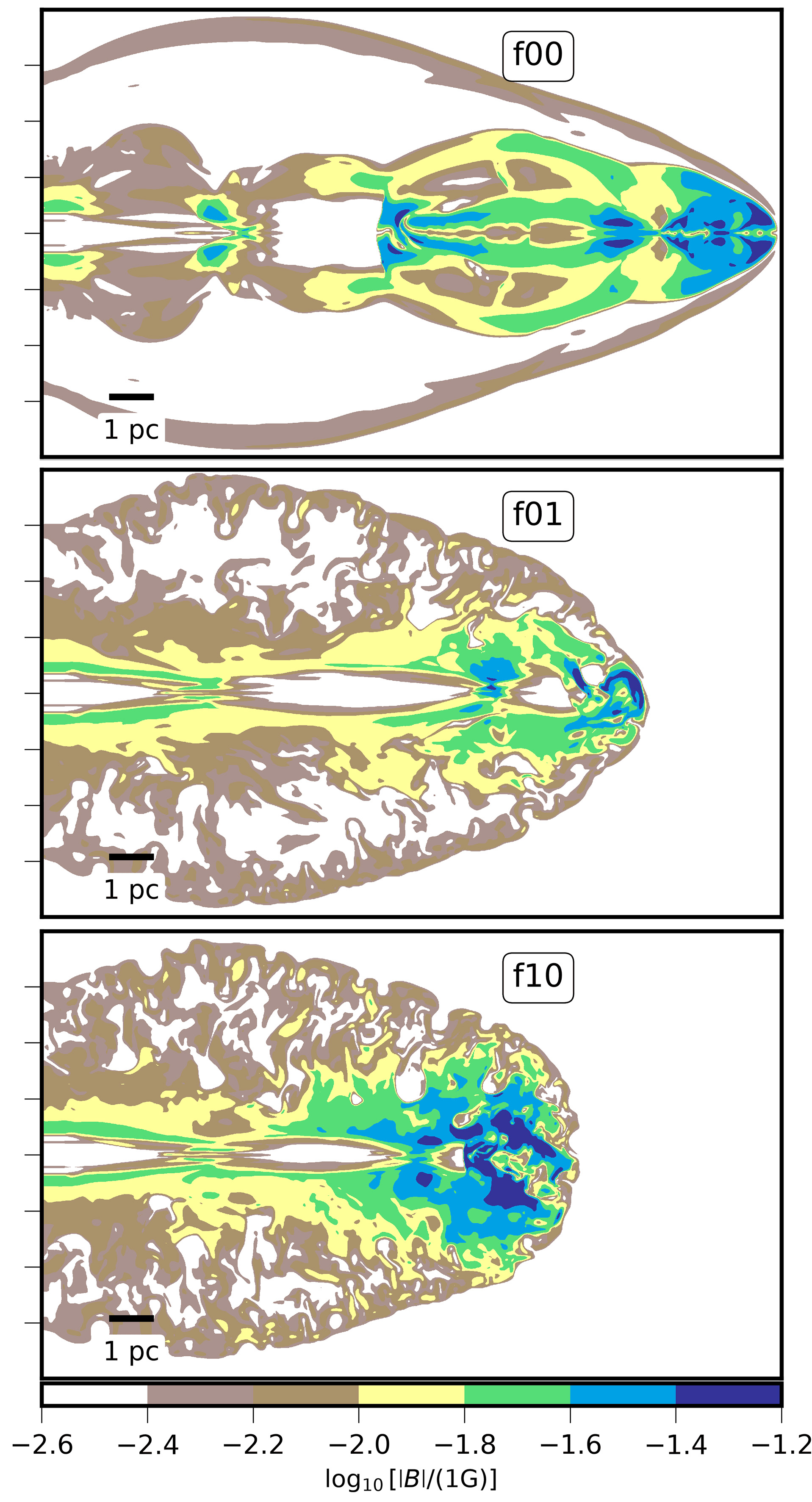}
\caption{
Magnetic-field magnitude, $|\mathbf{B}|$, at $t_{\rm sim}=85$~yr for \fZero\ (top), \fLow\ (middle), and \fHigh\ (bottom).
The field morphology is shaped by the ambient density structure and by compression in shocked regions.
}
\label{fig:jet-B}
\end{figure}

Figure~\ref{fig:jet-B} shows the magnetic-field magnitude, $|\mathbf{B}|$, for the same simulations shown in Figure~\ref{fig:jet-rho}. The magnetic-field morphology is shaped by the density structure of the ambient medium, but the largest values of $|\mathbf{B}|$ are associated with shocked and compressed regions. In the homogeneous case, \fZero, these regions are located along the jet spine (with some internal shocks), in the jet head, and the bow shock, which delineates the cocoon in both the density and magnetic-field maps.

In the homogeneous comparison runs, \fZeroB\ and \fZeroC, only the ambient density is increased in the initial conditions, while the initial magnetic field is kept unchanged. The larger values of $|\mathbf{B}|$ therefore develop during the simulation, mainly through enhanced compression in shocked regions. This effect is most evident at the bow shock near the jet head, where the denser ambient medium provides greater resistance to the jet propagation. 

The simulations with a clumpy medium produce a different magnetic morphology. Because the ambient medium is porous, the jet is locally deflected by dense clouds while part of the flow propagates through low-density channels. This generates a filamentary magnetic structure inside the cocoon, in contrast to the smoother morphology found in the homogeneous cases. The regions of highest $|\mathbf{B}|$ remain associated with strongly shocked gas, particularly close to the jet head, but additional local enhancements also occur where the jet interacts with dense clouds.

Overall, the strongest magnetic fields occur near the jet head and regions where the flow is deflected by dense clouds. These regions may provide favourable conditions for magnetic reconnection and non-thermal particle acceleration \citep{Sironi2014}, although the present simulations do not resolve these processes explicitly.

\begin{figure}
\includegraphics[width=.97\linewidth]{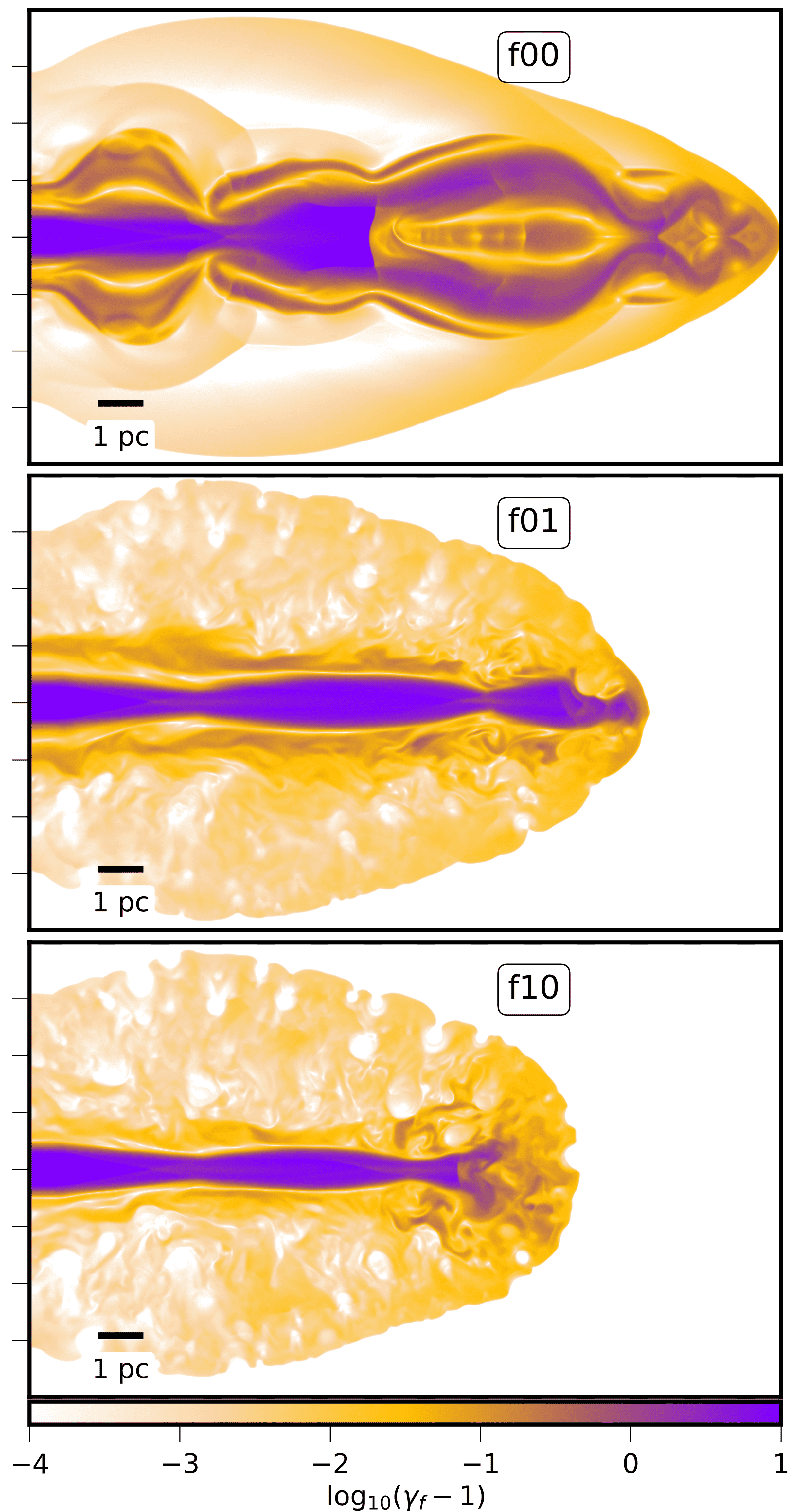}
\caption{
Fluid Lorentz factor maps, $\log_{10} \left( \gamma_f -1 \right)$, at $t_{\rm sim}=85$~yr for \fZero\ (top), \fLow\ (middle), and \fHigh\ (bottom).
This diagnostic highlights both the relativistic jet beam and the slower cocoon flow.
}
\label{fig:jet-u}
\end{figure}

To further characterise the flow structure, Figure~\ref{fig:jet-u} shows the fluid Lorentz factor maps, $\log_{10} \left( \gamma_f -1 \right)$, at $t_{\rm sim}=85$~yr. This quantity is a useful diagnostic because it retains contrast between mildly and highly relativistic material, while still showing velocity variations within the slower cocoon.

In the homogeneous run, the fast beam does not form a single coherent spine over the full jet length. Instead, the map shows an abrupt decrease of $\gamma_f$ along the jet axis, consistent with a strong internal shock associated with a large-scale backflow, as also discussed in \textit{D24}. 
In the simulations with a clumpy medium, the high-$\gamma_f$ material remains more continuous along the jet axis. Dense clouds perturb and deflect the flow, but they also suppress the development of the same coherent backflow structure seen in the homogeneous case. 
The clumpy ISM runs therefore show a more persistent fast channel, together with a broader cocoon of slower, stirred gas.

In the clumpy ISM simulations, the high-$\gamma_f$ channel also shows a sharp narrowing before the jet head, suggestive of an additional recollimation shock. 
The immediate jet-head region is complex, with multiple shocks and strong flow deflections, so individual shock features cannot be unambiguously separated using this diagnostic alone.

Overall, these maps show that the ambient density structure controls both the global propagation of the jet and the internal organisation of the flow. Increasing the total gas mass reduces the average jet-head advance speed, while clumpiness modifies how the jet couples to the surrounding medium by introducing local obstacles and low-density channels. The result is a less regular cocoon morphology, a more structured magnetic field, and a fast component that can remain channelled through the porous medium. These differences motivate the quantitative analysis of mass loading, entrainment, and mixing presented in the following sections.

\subsection{Mass Loading of the Jet-driven Outflow} \label{subsec:massloading}

We now quantify how the ambient medium affects the amount of material carried by the jet-driven flow. Motivated by the morphological differences discussed above, we separate the gas into jet material, entrained material, and unselected material as it follows:
\begin{itemize}[label={--}, leftmargin=*]
    \item \textit{Jet material:} cells with $v > 0.2\,c$ and $\sigma > 0.3$, corresponding to fast, highly magnetised gas;
    \item \textit{Entrained material:} cells with $10^{-3}\,c < v \leq 0.2\,c$ and $10^{-3} < \sigma \leq 0.3$;
    \item \textit{Unselected material:} all remaining cells that do not satisfy the previous conditions.
\end{itemize}
Here $\sigma \equiv B^2/\rho c^2$ is the magnetisation parameter. The entrained component therefore corresponds to slower, less magnetised gas that has been set in motion by the jet, rather than to all gas of ambient origin.
We verified that reasonable variations of these thresholds do not appreciably change the trends reported below.
This classification does not by itself determine the origin of the gas. The material origin is analysed separately using passive tracers in Section~\ref{sec:material-mixing}.
Using these definitions, the total jet mass is computed as
\begin{equation}
\begin{aligned}
M_{\rm jet}(t) &= \int_V \gamma \rho \, \mathcal{J}\, dV, \\ 
\mathcal{J} &=  \begin{cases}
    1 & \text{if } v > 0.2\,c \text{ and } \sigma > 0.3, \\
    0 & \text{otherwise}.
    \end{cases}
\end{aligned}
\label{eq:Mjet-def}
\end{equation}
where $\mathcal{J}$ is the jet selection function. Similarly, the entrained mass is
\begin{equation}
\begin{aligned}
M_{\rm ent}(t) &= \int_V \gamma \rho \, \mathcal{E}\,dV, \\
\mathcal{E} &=  \begin{cases}
    1 & \text{if } 10^{-3}\,c < v \leq 0.2\,c \\
    & \text{ and } 10^{-3} < \sigma \leq 0.3, \\
    0 & \text{otherwise}.
    \end{cases}
\end{aligned}
\label{eq:Ment-def}
\end{equation}
The ratio
\begin{equation}
\eta_M \equiv \frac{M_{\rm ent}(z = 20{\rm pc})}{M_{\rm jet}(z = 20{\rm pc})}
\label{eq:etaM-def}
\end{equation}
quantifies the degree of mass loading of the jet-driven outflow.

\begin{table}
\centering
\begin{tabular}{l|ccc|cccc}
&
\multicolumn{3}{c|}{$t_{\rm sim}=85$~yr} &
\multicolumn{4}{c}{$z_{\rm j}\approx11$--$12$~pc} \\ \hline \hline
\textbf{Sim ID} &
$M_{\rm jet}$ &
$M_{\rm ent}$ &
$\eta_M$ &
$M_{\rm jet}$ &
$M_{\rm ent}$ &
$\eta_M$ &
$t_{\rm sim}$ \\
&
[$M_\odot$] &
[$M_\odot$] &
&
[$M_\odot$] &
[$M_\odot$] &
&
[yr] \\
\hline\hline
\fZero\  & 0.23 & 0.43 & 1.9 & 0.18 & 0.32 & 1.8 & 57 \\
\fLow\   & 0.24 & 5.3  & 22  & 0.19 & 4.2  & 22  & 73 \\
\fHigh\  & 0.18 & 8.8  & 49  & 0.18 & 8.8  & 48  & 85 \\
\hline\hline
\fZeroB\ & 0.18 & 0.51 & 2.8 & 0.16 & 0.47 & 2.9 & 73 \\
\fZeroC\ & 0.14 & 1.1  & 7.9 & 0.14 & 1.1  & 7.9 & 85 \\
\hline \hline
\end{tabular}
\caption{
Jet and entrained masses measured using the dynamical selection criteria defined in Equations~\eqref{eq:Mjet-def} and \eqref{eq:Ment-def}.
The first three columns give the values at fixed time, $t_{\rm sim}=85$~yr.
The last four columns give the corresponding values for all simulations at a comparable jet length, $z_{\rm j}\approx11$--$12$~pc. The cumulative profiles of $M_{\rm jet}$ and $M_{\rm ent}$ are shown in Figure~\ref{fig:jet-mass-time}, and their simulation times are listed in the final column.
The mass-loading ratio is defined as $\eta_M\equiv M_{\rm ent}/M_{\rm jet}$.
}

\label{tab:masses-loading}
\end{table}

\begin{figure*}
\includegraphics[width=\linewidth]{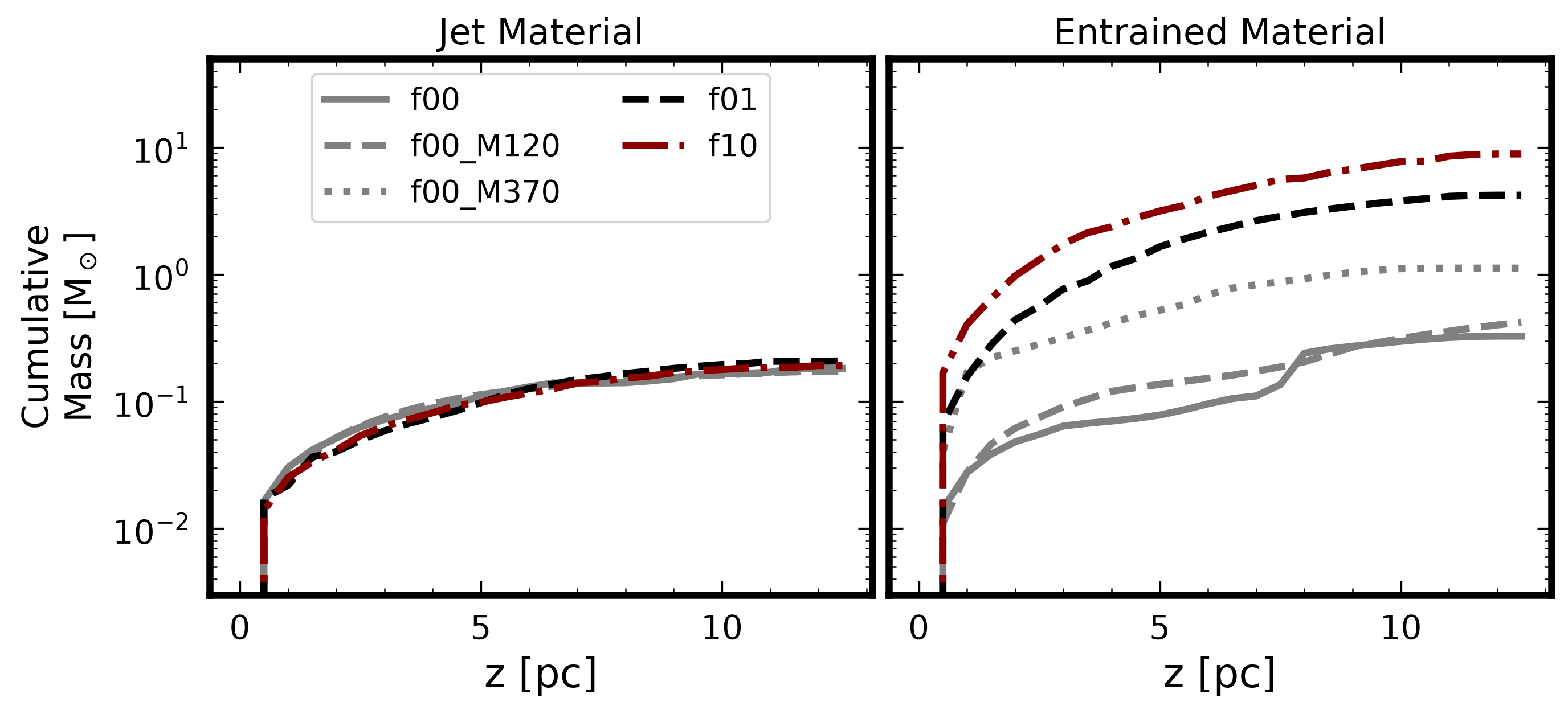}
\caption{
Cumulative jet mass (left) and entrained mass (right) as a function of distance along the jet axis, $z$, when the jets have reached comparable lengths, $z_{\rm j}\approx11$--$12$~pc. The corresponding simulation times are $57$, $73$, $85$, $73$, and $85$~yr for \fZero, \fLow, \fHigh, \fZeroB, and \fZeroC, respectively, and are also listed in Table~\ref{tab:masses-loading}. Lines indicate simulations \fZero\ (grey solid), \fZeroB\ (grey dashed), \fZeroC\ (grey dotted), \fLow\ (black dashed), and \fHigh\ (red dash-dotted). The grey dashed and dotted lines correspond to homogeneous comparison runs matched in total initial gas mass to \fLow\ and \fHigh, respectively.
}
\label{fig:jet-mass-time}
\end{figure*}

Figure~\ref{fig:jet-mass-time} shows the cumulative jet mass (left) and entrained mass (right) as a function of distance along the jet axis when the jets have reached comparable lengths, $z_{\rm j}\approx11$--$12$~pc. The corresponding simulation times are $57$, $73$, $85$, $73$, and $85$~yr for \fZero, \fLow, \fHigh, \fZeroB, and \fZeroC, respectively. 
The profiles are obtained by evaluating the volume integrals in Equations~\eqref{eq:Mjet-def} and \eqref{eq:Ment-def} cumulatively, including only cells with axial position smaller than a given $z$. The corresponding integrated values are listed in Table~\ref{tab:masses-loading}.

At comparable jet lengths, the mass of the fast, magnetised jet component varies only weakly among the three main simulations, with $M_{\rm jet}=0.18$--$0.19~{\rm M_\odot}$. 
The entrained component shows a markedly different trend. At comparable jet lengths, \fZero\ contains $M_{\rm ent}=0.32~{\rm M_\odot}$, whereas \fLow\ and \fHigh\ contain $4.2~{\rm M_\odot}$ and $8.8~{\rm M_\odot}$, respectively. The corresponding mass-loading ratios are $\eta_{\rm M}=1.8$, $22$, and $48$. Thus, when the simulations are compared at a similar propagation stage, the clumpy ISM models contain substantially more entrained material than the homogeneous model.

The homogeneous comparison simulations further show that this enhancement cannot be explained by the total gas mass alone. At fixed jet length, \fZeroB\ and \fZeroC\ contain $0.47~{\rm M_\odot}$ and $1.1~{\rm M_\odot}$ of entrained material, respectively, approximately an order of magnitude less than the corresponding clumpy simulations with matched initial gas masses. The spatial distribution of the ambient gas therefore plays a central role in determining the entrainment. In the clumpy simulations, low-density channels allow part of the jet to propagate through the porous ISM, while dense clouds are shocked, displaced, and partially disrupted into the cocoon, substantially increasing the amount of ambient material incorporated into the jet-driven outflow.

For completeness, Table~\ref{tab:masses-loading} also reports the integrated masses at the common simulation time $t_{\rm sim}=85$~yr. The fixed-time comparison yields the same overall trends, although it shows modest differences in $M_{\rm jet}$. This result mainly reflect the different distances reached by the jets during the simulation time.

In summary, the presence of a clumpy medium does not strongly alter the mass of fast, magnetised jet material selected by our criteria. It does, however, generate a much more massive, slower, and weakly magnetised component, which we identify as entrained material. Jet--cloud interactions shock, displace, and partially disrupt dense gas, increasing the amount of material carried by the cocoon, particularly towards the jet head. This entrained component dominates the outflow mass budget and is associated with the slower propagation and broader morphology found in the multiphase simulations.

\section{Material Mixing}\label{sec:material-mixing}

\subsection{Tracer Distributions}

The mass-loading analysis in Section~\ref{subsec:massloading} separates the flow into jet, entrained, and ambient components using kinematic and magnetic criteria. 
This classification identifies dynamically distinct regions, but does not determine the origin of the gas within them. 
To examine how injected jet material and ambient gas are distributed through the jet-driven flow, we use the passive scalar tracers evolved in the PLUTO simulations \citep{Mignone2007, Mignone2011}.

We define two passive tracers, $\mathcal{T}_{\rm j}$ and $\mathcal{T}_{\rm cl}$. 
The jet tracer, $\mathcal{T}_{\rm j}$, is initialised to unity inside the injection nozzle and traces material originally injected by the jet. 
The cloud tracer, $\mathcal{T}_{\rm cl}$, is initialised to unity inside the dense clouds and traces material initially belonging to the dense phase of the clumpy ISM ($n > 1$~cm$^{-3}$). 
Both tracers are initially zero elsewhere and are subsequently advected with the fluid. 
No independent tracer is assigned to the diffuse background medium. 
The tracers should therefore be interpreted as indicators of material origin, rather than as distinct particle populations or as the non-thermal macro-particles used to compute the synchrotron emission in Section~\ref{sec:synchrotron}.

\begin{figure*}
\includegraphics[width=\linewidth]{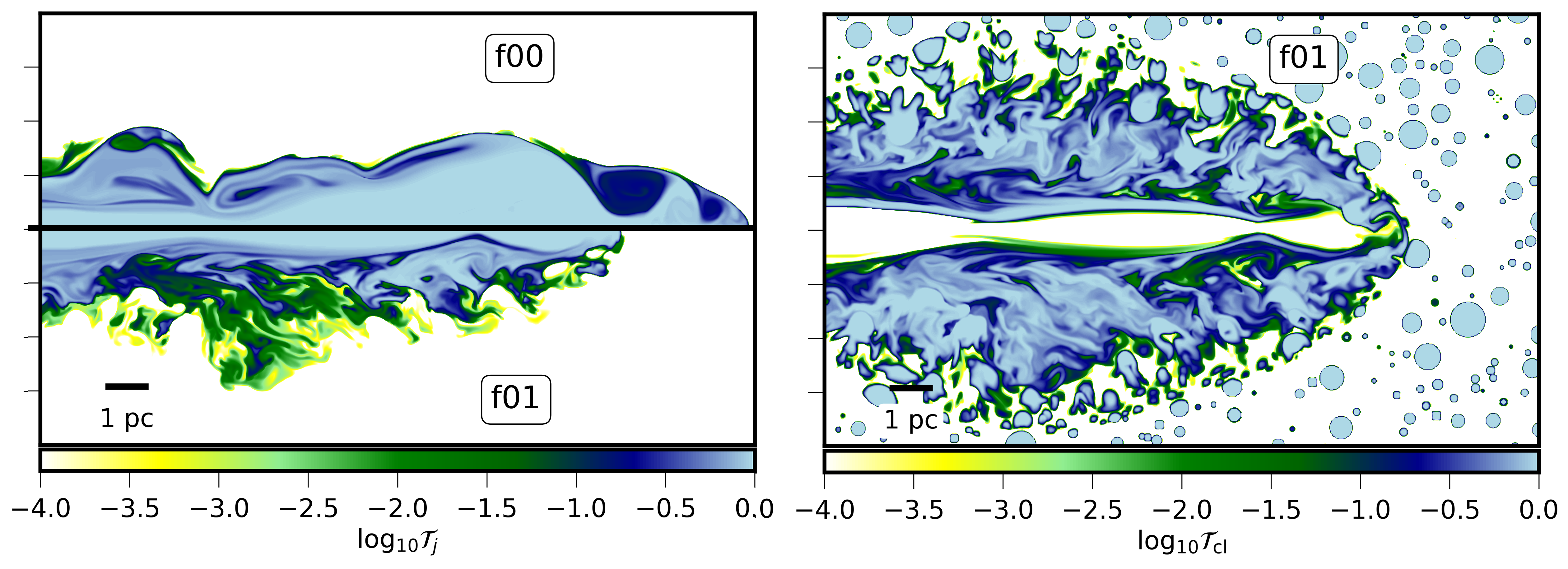}
\caption{
Distribution of the passive scalar tracers at $t=85$~yr associated with jet-origin material, $\mathcal{T}_{\rm j}$ (left), and cloud-origin material, $\mathcal{T}_{\rm cl}$ (right). 
\textit{Jet-origin material:} In the homogeneous case (top), the jet tracer closely follows the density map shown in Figure~\ref{fig:jet-rho}. 
In contrast, the simulation with a clumpy background shows a narrower distribution, with jet-origin material remaining concentrated near the central channel in the presence of dense clouds.
\textit{Cloud-origin material:} The cloud tracer shows that disrupted cloud-origin material is distributed throughout much of the cocoon, whereas the central jet-dominated channel remains primarily occupied by jet-origin material.
}
\label{fig:jet-tracer}
\end{figure*}

Figure~\ref{fig:jet-tracer} shows the distributions of $\mathcal{T}_{\rm j}$ and $\mathcal{T}_{\rm cl}$ at $t=85$~yr. 
In the homogeneous simulation (top left panel), the jet-origin material spreads through a broad cocoon and closely follows the jet morphology shown in Figure~\ref{fig:jet-rho}. In the simulation with a clumpy medium (bottom left panel), the jet-origin material remains concentrated within a narrower central channel, while lower tracer values extend into the surrounding cocoon. This distribution is consistent with the confinement and deflection of the jet by dense obstacles, together with the formation of a broad region containing diluted jet material, consistent with the velocity structure shown in Figure~\ref{fig:jet-u}.

The cloud tracer (right panel of Figure~\ref{fig:jet-tracer}) indicates that the initially spherical clouds are efficiently disrupted and redistributed throughout much of the cocoon volume, while remaining largely absent from the jet-dominated central channel.
Over much of the cocoon region, $\mathcal{T}_{\rm cl} \geq 0.5$, indicating that cloud-origin material locally dominates the gas composition in these regions.
At the same time, typical cocoon values of $\mathcal{T}_{\rm j} \approx 10^{-3}$ indicate that a small amount of jet-origin material is also present. 
The overlap between non-zero $\mathcal{T}_{\rm j}$ and $\mathcal{T}_{\rm cl}$ therefore indicates local coexistence, and hence mixing, between jet-origin and cloud-origin material within parts of the cocoon.
This tracer analysis complements the dynamical mass-loading diagnostics by showing that the entrained component contains gas of different origins. In the following subsection, we quantify the mixing between jet-origin material and the total ISM-origin contribution, where the latter includes both diffuse background gas and cloud-origin material.

\subsection{Mass-weighted Jet--ISM Mixing}\label{subsec:mass-weighted-mixing}

The tracer distributions in Figure~\ref{fig:jet-tracer} show that material of different origins is present within the jet-driven flow. We now quantify the mixing between jet-origin and ISM-origin material separately within the jet and entrained components defined in Section~\ref{subsec:massloading}. The jet component is selected by $\mathcal{J}$ (Equation~\eqref{eq:Mjet-def}), corresponding to fast and highly magnetised gas, whereas the entrained component is selected by $\mathcal{E}$ (Equation~\eqref{eq:Ment-def}), corresponding to slower, less magnetised outflowing gas that does not satisfy the jet criteria.

For this analysis, the ISM-origin contribution is defined as $\mathcal{T}_{\rm ISM}=1-\mathcal{T}_{\rm j}$, and therefore includes both diffuse background material and, in the simulations considering a clumpy medium, gas initially associated with dense clouds. This quantity is distinct from $\mathcal{T}_{\rm cl}$, which traces only cloud-origin material.
Following \citet{Walg2013}, we define a mass-weighted mixing factor
\begin{equation}
\Lambda = 1 - \left| \frac{\mathcal{T}_{\rm j} -  \mu_{\rm M}\mathcal{T}_{\rm ISM}}{\mathcal{T}_{\rm j} +  \mu_{\rm M}\mathcal{T}_{\rm ISM}}\right| 
= 1 - \left| \frac{(1 + \mu_{\rm M}) \mathcal{T}_{\rm j} - \mu_{\rm M}} {(1- \mu_{\rm M}) \mathcal{T}_{\rm j} + \mu_{\rm M}} \right|.
\label{eq:mass-mixing-factor}
\end{equation}
Here, $\mu_{\rm M}$ is the ratio between the total masses of jet-origin and ISM-origin material within the combined jet and entrained components, it accounting
for the unequal masses of jet-origin and ISM-origin material. 
To compare mixing within the jet and entrained components using the same reference mixture, we calculate $\mu_{\rm M}$ over the combined dynamically selected jet 
and entrained components,
\begin{equation}
\mu_{\rm M} 
 = \frac{
\displaystyle \int
\gamma \rho \mathcal{T}_{\rm j}
\left(\mathcal{J}+\mathcal{E}\right) \, {\rm d}V
}{
\displaystyle \int
\gamma \rho (1 - \mathcal{T}_{\rm j})
\left(\mathcal{J}+\mathcal{E}\right) \, {\rm d}V
} .
\label{eq:mu-M}
\end{equation}
For the combined dynamically selected jet and entrained components, the mass ratios between jet-origin and ISM-origin material are $\mu_{\rm M}=3.9 \times 10^{-3}$, $1.5 \times 10^{-3}$, and $5.7 \times 10^{-4}$ for \fZero, \fLow, and \fHigh, respectively.

With this definition, $\Lambda=0$ corresponds to locally unmixed material, while $\Lambda=1$ indicates that the local jet-to-ISM mass ratio is equal to that of the dynamically selected jet-driven material as a whole. 

Since $\Lambda$ is defined locally, we characterise the mixing within each dynamical component by averaging its value over all cells selected as jet or entrained material. As all computational cells have equal volume, these arithmetic means correspond to volume-averaged mixing factors, denoted by $\left\langle\Lambda\right\rangle_{\rm jet}$ and $\left\langle\Lambda\right\rangle_{\rm ent}$, respectively.

At $t=85$~yr, the volume-averaged mixing factor of the entrained component is $\left\langle\Lambda\right\rangle_{\rm ent}=1.1$, $5.7$, and $7.9$ per cent for \fZero, \fLow, and \fHigh, respectively, with corresponding standard deviations of $7.8$, $17.2$, and $20.0$ per cent. For the jet component, the corresponding values are $\left\langle\Lambda\right\rangle_{\rm jet}=2.3$, $4.2$, and $8.9$ per cent, with standard deviations of $9.4$, $13.5$, and $19.9$ per cent. Thus, increasing the cloud filling factor systematically increases the average jet--ISM mixing in both dynamically selected components. However, the standard deviations substantially exceed the corresponding mean values, particularly in the clumpy ISM simulations, showing that the mixing is highly inhomogeneous. The disturbed cocoon morphology and disrupted cloud structures therefore do not correspond to uniform compositional homogenisation: instead, enhanced mixing is confined to a subset of the selected cells, while much of the material remains dominated by either jet-origin or ISM-origin gas.

The low average values of $\Lambda$ should therefore not be interpreted as evidence for weak dynamical coupling between the jet and the ISM. The density morphology, cloud disruption, and turbulent density distributions (see Section~\ref{sec:turbulence}) demonstrate that the clumpy medium is strongly processed by the jet. Rather, $\Lambda$ measures the more restrictive condition of local homogenisation between jet-origin and total ISM-origin material, which remains incomplete on the simulated time-scale.

\subsection{Momentum Transfer and Jet--ISM Coupling}
\label{subsec:jet-ism-coupling}

Figure~\ref{fig:jet-mass-time} shows the total jet mass remains essentially unchanged across our simulations, as it is primarily set by the injection conditions at the nozzle. In contrast, the entrained mass increases with the clump volume filling factor, indicating that mass loading is controlled not only by the amount of gas available in the domain, but also by how efficiently the jet couples to the surrounding medium. The spatial distribution of the ISM is therefore central to the evolution of the outflow \citep{Bicknell2003, Ward2024}.

In the homogeneous case, the jet propagates more coherently and entrains gas mainly through boundary layers along the jet and cocoon. In the simulations considering a clumpy ISM, dense structures intercept the flow, increasing the effective interaction surface between the jet and the ISM. This produces repeated deflections, shocks, and shear layers, which drive turbulence and promote the stripping of dense gas from the clouds \citep{Sutherland2007, Wagner2012, Wagner2013, Mukherjee2018}. Low-density regions between clumps allow the jet to penetrate the medium while continuing to ablate nearby dense structures. These processes collectively increase the amount of ambient material incorporated into the jet-driven outflow.

Using $\mathcal{T}_{\rm j}$ and $\mathcal{T}_{\rm ISM}$, we calculate the tracer-weighted matter contribution to the axial momentum carried by material of different origins over the computational domain. For each constituent $k \in \left\{ {\rm jet}, {\rm ISM} \right\}$, we define

\begin{equation}
P_{z,k} =
\int \gamma^2 \rho h v_z \mathcal{T}_{k} \, {\rm d}V,
\label{eq:tracer-axial-momentum}
\end{equation}
Material in the unperturbed ambient medium makes no contribution to this quantity because its axial velocity is zero. The integral therefore predominantly measures the matter axial momentum carried by jet-origin and ISM-origin material participating in the jet-driven flow.

At $t=85$~yr, the jet-origin component carries an axial momentum of $P_{z,{\rm j}} \simeq 2.5$--$3.5 \times 10^{5}~{\rm M_\odot \, km \, s^{-1}}$, whereas the ISM-origin component carries $P_{z,{\rm ISM}} \simeq 1.5$--$2.5 \times 10^{4}~{\rm M_\odot \, km \, s^{-1}}$. The matter axial momentum budget therefore remains dominated by material injected through the jet nozzle. In the clumpy ISM simulations, ISM-origin material contributes only $\sim 5$--$10$ per cent of the combined jet--ISM axial momentum, despite representing a substantial fraction of the entrained mass.

Although the tracer maps show that cloud material is efficiently stripped and incorporated into the cocoon, the momentum analysis indicates that most of the net matter axial momentum remains associated with injected jet-origin material.
The jet disrupts the clouds but is unable to accelerate them efficiently within the analysed time-scale. At the same time, jet-origin and total ISM-origin material do not become fully homogenised. Cloud stripping, jet--ISM mixing, and cloud acceleration therefore do not proceed with equal efficiency.

Clouds interacting with an external flow, the jet in our setup, have a characteristic cloud-crushing timescale,
\begin{equation}
    t_{\rm cc} = \chi^{1/2} \frac{r_{\rm cl}}{v_{\rm jet}},
\end{equation}
and an acceleration timescale,
\begin{equation}
    t_{\rm acc} = \chi^{1/2} t_{\rm cc},
\end{equation}
where $\chi = \rho_{\rm cl}/\rho_{\rm jet}$ is the density contrast and $r_{\rm cl}$ is the cloud radius \citep{Klein1994, Gronke2018, Ji2019, Fielding2020, Dutta2025}.
In the present context, $t_{\rm cc}$ provides an approximate timescale for the disruption and incorporation of cloud material into the cocoon (as seen in the right panel of Figure~\ref{fig:jet-tracer}), while $t_{\rm acc}$ estimates the time required for this material to be accelerated by the jet. 

For our typical clouds, $t_{\rm cc} \sim 50-200$~yr and $t_{\rm acc} \gtrsim 1500$~yr. The smaller clouds can therefore be disrupted into the cocoon within the simulated time, but they are not expected to be accelerated to velocities comparable to the jet. This is consistent with the tracer-weighted momentum budget: 
cloud-origin material locally dominates much of the cocoon, but remains dynamically sub-dominant in terms of net matter axial momentum.

The main dynamical effect of the clumpy medium at the scales probed here is therefore to increase the inertia of the jet-driven outflow, enhance mass loading, and contribute to jet deflection and deceleration, rather than to transfer a large fraction of the jet momentum to the dense gas. The difference between homogeneous and clumpy environments arises not from the total available gas mass alone, but from how that mass is structured. Even modest filling factors can substantially increase the interaction between the jet and the ISM, producing more irregular morphologies, stronger turbulence, and a more massive entrained component \citep{Bieri2016, Ward2024}.

The altered jet--ISM coupling in the clumpy ISM simulations is also expected to influence the non-thermal emission. Enhanced jet deflection, cloud disruption, and the development of a more structured cocoon modify the distribution of shocks, shear layers, and regions of compressed magnetic field through which synchrotron-emitting particles evolve. The resulting emission morphology need not trace the mass distribution directly: although the entrained component dominates much of the cocoon mass, the synchrotron signal is determined by the spatial distribution and energy losses of the non-thermal particle population. In the following section, we examine how these dynamical differences are imprinted on the synthetic synchrotron emission maps.

\section{Synchrotron Signatures} \label{sec:synchrotron}
\begin{figure*}
\includegraphics[width=\linewidth]{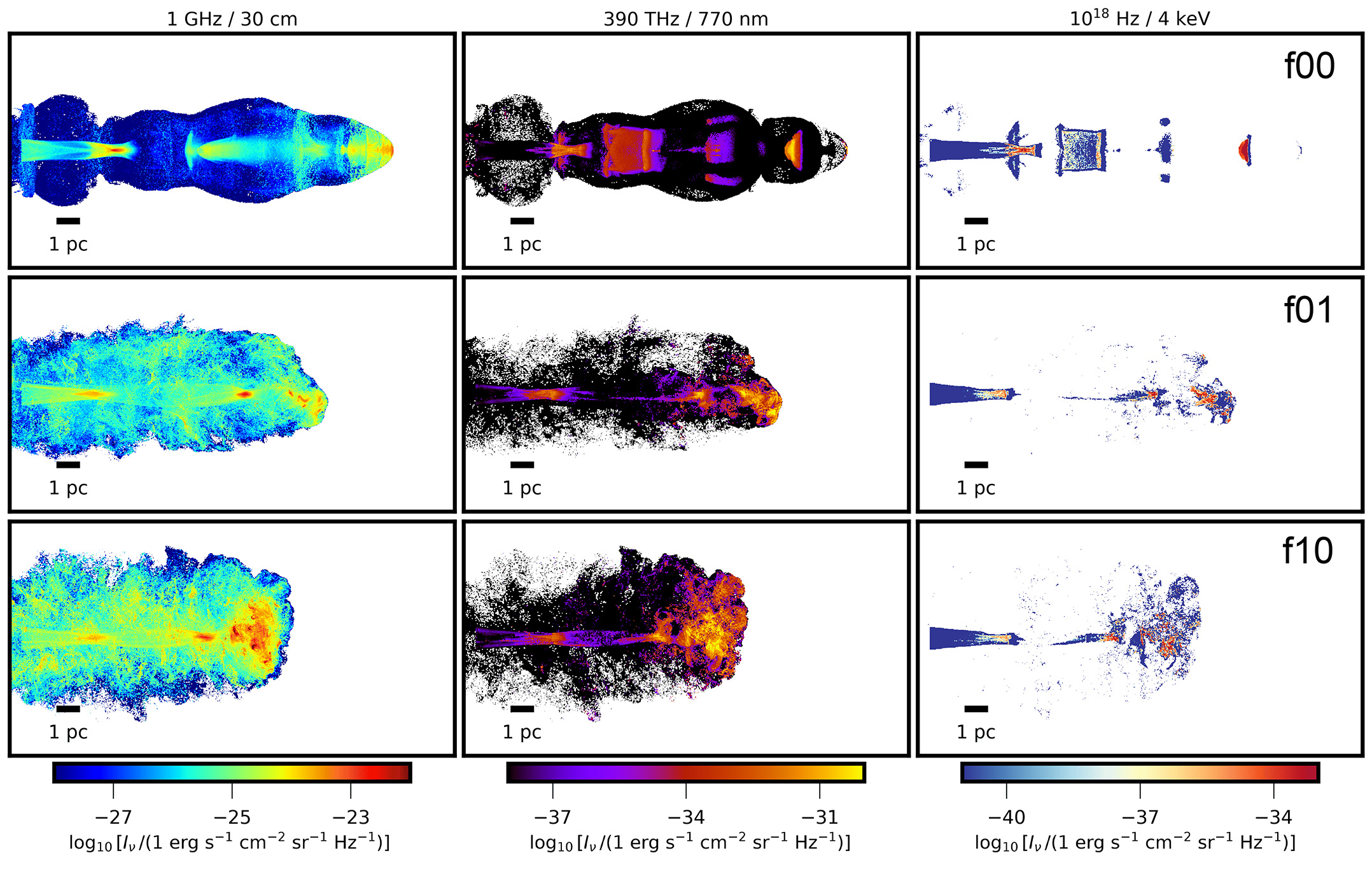}
\caption{Synchrotron specific intensity $I_\nu$ at $\nu = 1$~GHz (left), $390$~THz (centre), and $4$~keV (right) for the three main simulations at $t = 85$~yr, assuming a viewing angle of $90^\circ$. In the homogeneous case, the emission is dominated by the jet spine and terminal shock, producing an elongated and symmetric morphology. In contrast, the presence of a multiphase ISM broadens and fragments the emission, with multiple localised bright regions associated with jet--cloud interactions and shock compression. The \fHigh\ simulation shows the strongest emission, concentrated near the jet head, consistent with enhanced particle acceleration in a dense, clumpy environment.}
\label{fig:EM-Inu}
\end{figure*}
To connect the dynamical evolution discussed above with observable non-thermal signatures, we analyse the synchrotron emission using the Lagrangian macro-particles injected with the jet (Section~\ref{sec:model}). 
The emitting particle population evolves through the jet and cocoon, where its energy distribution is shaped by adiabatic, synchrotron, and inverse-Compton losses, as well as by shock acceleration. 
Since the emissivity also depends on the local magnetic-field strength, the resulting maps provide a diagnostic of how jet propagation, magnetic-field amplification, and jet--ISM interactions modify the apparent morphology of the source. 
In this section, we compare the frequency-dependent synchrotron morphology and integrated spectra of the homogeneous and clumpy simulations.

Figure~\ref{fig:EM-Inu} shows the synchrotron specific intensity, $I_{\nu}$, at $t=85\,{\rm yr}$ for the three ambient-medium configurations, evaluated at representative radio, optical/near-infrared, and X-ray frequencies. 
These maps illustrate how the same underlying jet--ISM interaction produces distinct emission morphologies across the spectrum. 
At low frequencies, the emission traces the extended jet and cocoon structure, whereas at higher frequencies it becomes increasingly concentrated around compact regions where energetic particles and amplified magnetic fields coexist. 
The comparison between the homogeneous and clumpy simulations therefore provides a frequency-dependent view of how ISM structure affects the apparent morphology of young radio sources.

\subsection{Radio signatures}\label{subsec:radio}

At $1$~GHz, the synchrotron emission broadly traces the extended jet--cocoon morphology. 
In the \fZero\ simulation, the emission follows the jet structure seen in Figures~\ref{fig:jet-rho} and~\ref{fig:jet-B}, with bright regions along the jet spine and near the jet head, where shocks are present. 
In the clumpy ISM simulations, the radio-emitting region becomes wider and more irregular, reflecting the stronger distortion of the cocoon and the broader distribution of emitting particles. 
The cocoon contribution is enhanced relative to \fZero, especially in \fHigh, where the interaction with dense clouds produces a more fragmented morphology and brighter emission around the jet head. 
In all cases, the radio intensity peaks near the terminal region, where shocks, magnetic-field amplification, and recently accelerated particles coexist. 
The low-frequency emission therefore provides a useful diagnostic of the global extent and morphology of the jet-driven outflow, although it also depends on the distribution of non-thermal particles and magnetic fields.

\subsection{Optical/near-infrared emission}\label{subsec:nir}

At $390$~THz, corresponding to $\lambda \simeq 770$~nm, the synchrotron emission becomes more spatially concentrated than in the radio band. 
In the \fZero\ simulation, the emission is dominated by the jet spine, with several compact knots along the jet and an extended emitting region near the jet head. 
In the clumpy-medium simulations, the emission is more confined to the inner jet channel and terminal region, with a weaker contribution from the extended cocoon. 
This indicates that the higher-frequency emission preferentially traces regions where energetic particles and stronger magnetic fields coexist, rather than the full extent of the cocoon. 
The \fHigh\ simulation is the brightest case at this frequency, consistent with enhanced magnetic-field amplification and particle energisation in the strongly disturbed jet--cloud interaction region.

\subsection{X-ray emission}\label{subsec:x-rays}

At $4$~keV, the synchrotron emission is confined to compact regions within the jet and near the jet head. 
This reflects the rapid radiative cooling of the highest-energy electrons, which limits the spatial extent of the X-ray-emitting population compared with the radio-emitting particles.
In the \fZero\ simulation, the emission is concentrated around the strongest shocks within the jet. 
In the clumpy ISM simulations, the X-ray morphology becomes more fragmented, with compact emitting regions associated with jet deflection, cloud interaction, and the disrupted jet-head structure. 
The \fHigh\ case shows the brightest and most irregular X-ray emission, consistent with stronger magnetic-field amplification and localised particle energisation in the jet--cloud interaction region. 
Thus, while the radio emission traces the broader jet--cocoon structure, the X-ray synchrotron emission highlights the most compact and energetic regions of the simulated source.

\subsection{Integrated synchrotron spectra}\label{subsec:sed}

\begin{figure}
\includegraphics[width=\linewidth]{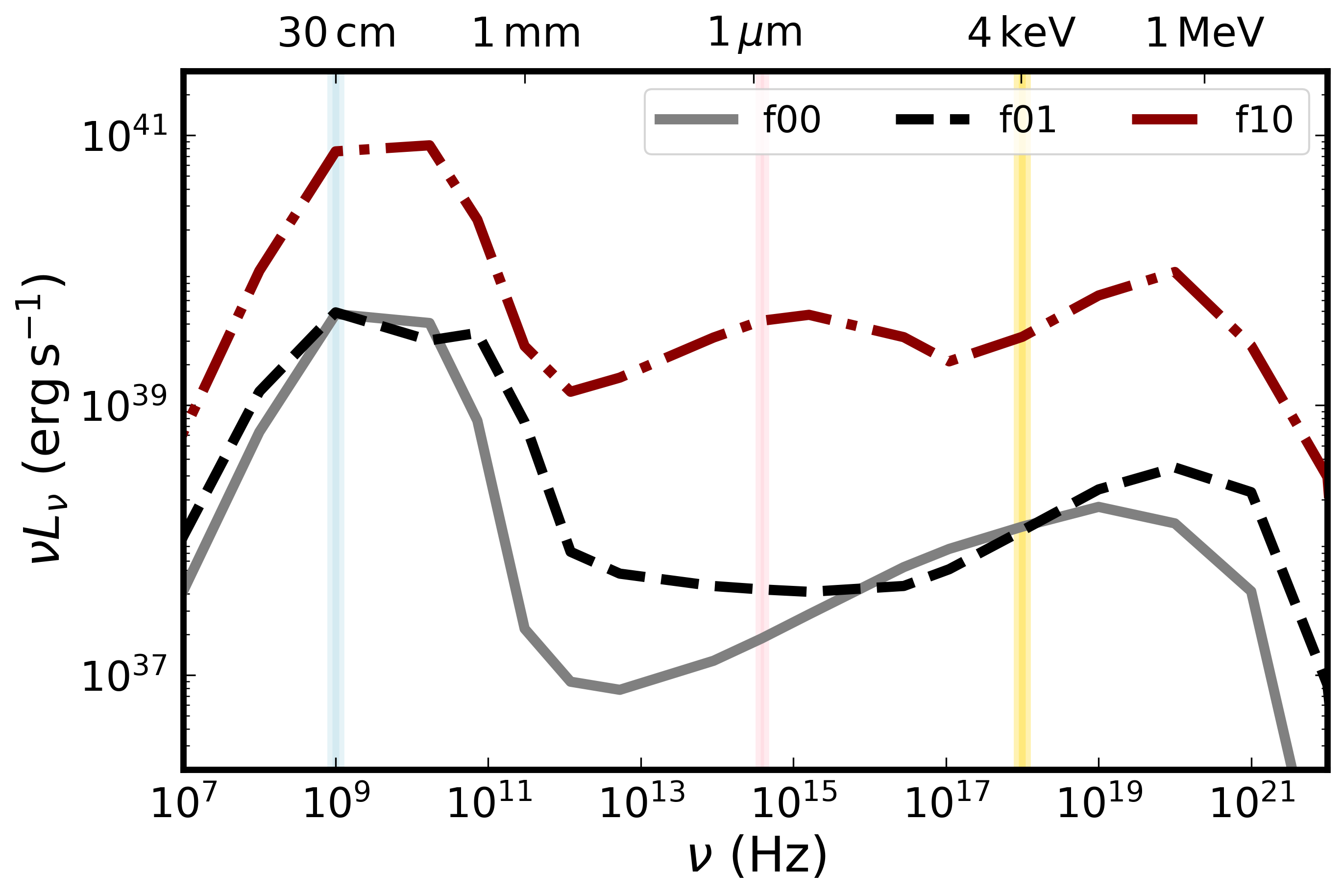}
\caption{
Integrated synchrotron spectra at $t = 85$~yr for the three simulations with different clump volume filling factors: \fZero\ (solid grey), \fLow\ (dashed black), and \fHigh\ (dash--dotted red). 
The vertical lines mark the three representative frequencies analysed in Figure~\ref{fig:EM-Inu}: $1$~GHz, $390$~THz, and $4$~keV.} 
\label{fig:SED}
\end{figure}

Figure~\ref{fig:SED} shows the integrated synchrotron spectra of the three simulations at $t=85\,{\rm yr}$. 
The spectra are shown as $\nu L_{\nu}$, and therefore indicate the frequency range in which most of the synchrotron power is emitted. 
All simulations show a prominent radio peak at $\nu \sim 10^9$--$10^{10}\,{\rm Hz}$ and a second broad component at high energies. 
The total luminosity increases with the clump volume filling factor, with \fHigh\ being systematically brighter than both \fZero\ and \fLow.

For the adopted dimensional scaling, the total injected jet power is 
$P_{\rm jet}=4.3\times10^{44}\,{\rm erg\,s^{-1}}$. 
The frequency-integrated synchrotron luminosities are 
$L_{\rm syn}=6.3\times10^{41}\,{\rm erg\,s^{-1}}$, $7.7\times10^{41}\,{\rm erg\,s^{-1}}$, and 
$1.3\times10^{43}\,{\rm erg\,s^{-1}}$ for \fZero, \fLow, and \fHigh, respectively. 
These correspond to ratios of 
$L_{\rm syn}/P_{\rm jet}\simeq 0.15$~per cent, 
$0.18$~per cent, and $3.0$~per cent. 
Thus, even in the brightest clumpy case, the integrated synchrotron emission remains only a few per cent of the injected jet power.
These efficiencies should be interpreted as post-processed synchrotron outputs rather than fully self-consistent radiative efficiencies, since the non-thermal particles are passive and their emission does not remove energy from the flow.

The homogeneous simulation shows the double-peaked synchrotron spectrum discussed in \textit{D24}, with a radio peak and a high-energy component. 
The spectral shape is governed by the evolved non-thermal particle distribution with multiple populations of electrons, the magnetic-field structure, and the radiative and adiabatic losses experienced by the emitting particles.

The spectral differences are not a simple monotonic rescaling of the homogeneous case. 
At radio frequencies, the \fLow\ spectrum remains close to \fZero, indicating that a low clump filling factor does not strongly modify the dominant low-frequency synchrotron component. 
The main difference appears at intermediate frequencies, where \fLow\ shows an excess relative to \fZero\ in the infrared/optical range. 
A smaller enhancement is also present at high energies, suggesting that modest clumpiness can slightly increase the contribution from the highest-energy emitting particles, although this effect is weaker than the infrared/optical excess.

The \fHigh\ simulation, by contrast, is substantially brighter across the full frequency range. 
The enhancement is particularly pronounced at infrared/optical frequencies, where the spectrum develops a broad excess between the radio and high-energy components. 
This behaviour is consistent with the stronger disruption of the large-scale jet flow discussed in Section~\ref{subsec:jet-structure}: the interaction with a larger number of dense clouds fragments the jet head and cocoon, producing a more extended distribution of localised shocks and shear regions. 
These sites can increase the contribution from intermediate-energy particles, while the strongest shocks continue to shape the high-energy synchrotron component.

The enhancement at infrared/optical frequencies can therefore be understood as a consequence of the different distribution of particle-acceleration sites. 
In the homogeneous simulation, the emission is more closely associated with coherent structures within the jet spine, including recollimation shocks, internal shocks, and the jet-head shock region, as also discussed in \textit{D24}. 
By contrast, in the clumpy simulations, jet deflection, cloud disruption, and shear at the jet--cloud interface generate a larger number of localised shocks and shear regions. 
These regions can populate the intermediate-energy part of the electron distribution, producing the enhanced infrared/optical emission seen in \fLow\ and, more prominently, in \fHigh. 
The high-energy component is only modestly enhanced in \fLow, but increases substantially in \fHigh, where the stronger jet--cloud interaction produces a more disrupted and shock-rich regions.

\subsection{Mock synchrotron maps}\label{subsec:mock-synchrotron-maps}

\begin{figure*}
\includegraphics[width=\linewidth]{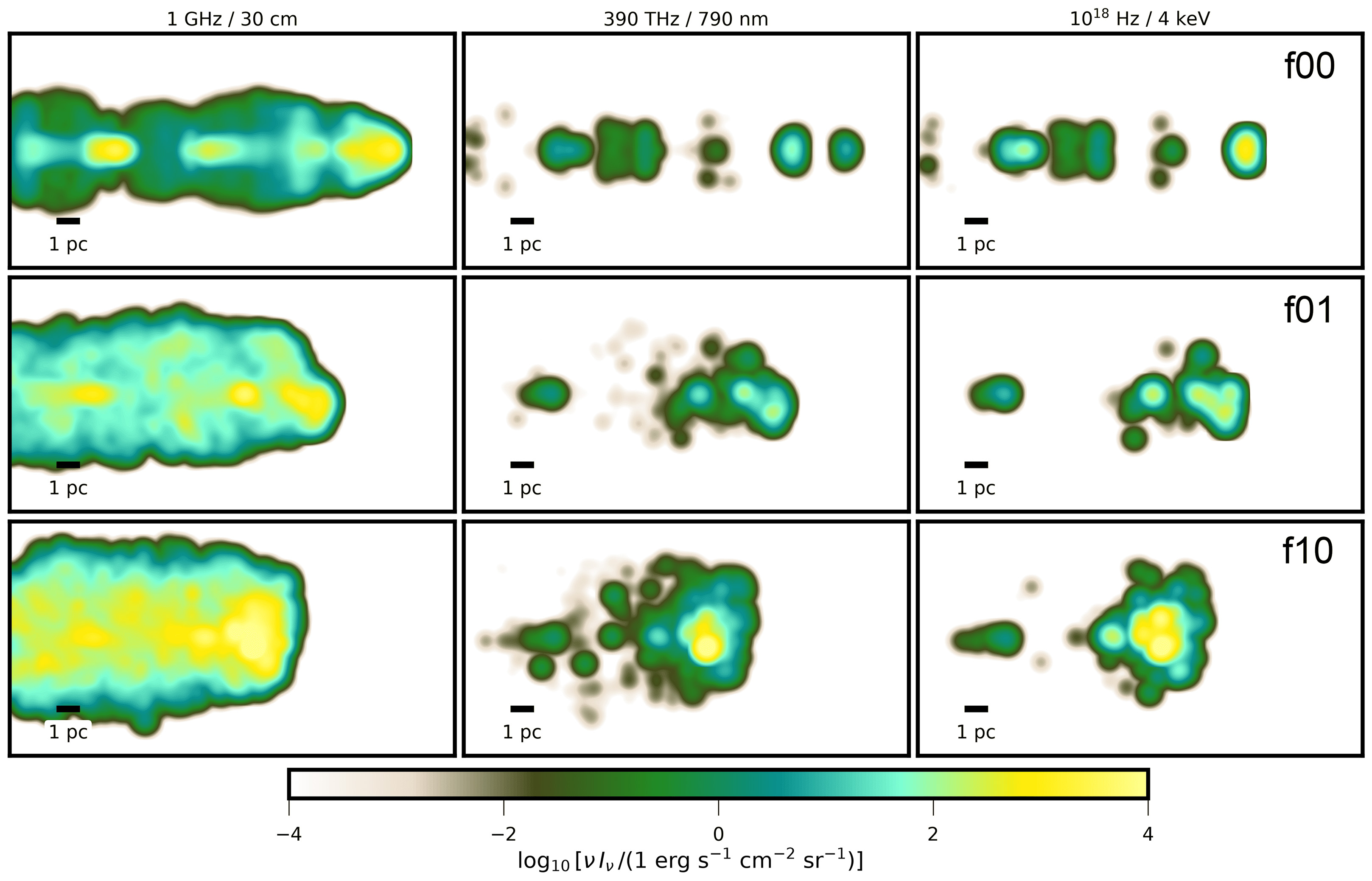}
\caption{
Mock synchrotron surface brightness maps for the three simulations at $t = 85$~yr. 
Rows correspond to different clump volume filling factors: \fZero\ (top), \fLow\ (middle), and \fHigh\ (bottom). 
Columns show three representative observing frequencies: $\nu = 1$~GHz, $390$~THz, and $4$~keV, from left to right. 
The maps are obtained by integrating the synchrotron emission along the line of sight for an inclination angle of $90^\circ$ and convolving the result with a Gaussian beam of width $0.2$~pc. 
The colour scale shows the logarithm of the surface brightness. 
}
\label{fig:EM-map}
\end{figure*}

In addition to the frequency-dependent synchrotron maps analysed in Sections~\ref{subsec:radio}--\ref{subsec:x-rays}, Figure~\ref{fig:EM-map} shows mock synchrotron surface-brightness maps constructed from the same particle-emission data as Figure~\ref{fig:EM-Inu}. 
For each model, we integrate the emission along the line of sight assuming an inclination angle of $90^\circ$ and convolve the resulting maps with a Gaussian beam of width $0.2$~pc, representing a circular telescope beam, as done in \textit{D24}. 
This procedure smooths compact structures and blends neighbouring emitting regions, providing a closer approximation to how the simulated sources would appear in finite-resolution observations. 
The maps are shown on a logarithmic scale for three representative observing bands: radio ($1$~GHz), optical/near-infrared ($390$~THz), and X-ray ($4$~keV). 
In producing these mock observations, we assume that the medium between the simulated jet and the observer is optically thin and neglect light-travel-time effects.

Despite the smoothing, the main morphological differences between the homogeneous and clumpy simulations remain visible. 
The homogeneous run, \fZero, appears relatively elongated and coherent, whereas the clumpy runs produce broader and more irregular projected structures. 
This effect is strongest in \fHigh, where the emission is laterally extended around the disrupted jet head and neighbouring bright regions are blended into a broader lobe structure. 
At higher frequencies, the emission remains more compact, but the clumpy cases still show fragmented and displaced bright regions near the jet head, consistent with the more complex shock structure produced by jet--cloud interactions.
These differences indicate that observed jet morphology may reflect not only intrinsic jet properties, but also the structure of the ambient medium.

\subsection{Implications for compact radio sources}\label{subsec:compact-radio-sources}

Our simulation box covers scales of order $\sim 50$~pc, whereas most observed radio jets are detected on larger scales because of their age and distance. 
A particularly relevant class of objects is compact symmetric objects (CSOs), radio sources smaller than $1$~kpc that show emission on both sides of their centre of activity and are not dominated by relativistic beaming towards the observer \citep{Wilkinson1994,Readhead1996}. 
These systems may be young radio sources or objects undergoing short episodes of activity, with lifetimes shorter than a few thousand years \citep{Kiehlmann2024a,Kiehlmann2024b,Readhead2024}. 
CSOs show a wide range of morphologies: some have well-defined lobes, such as 1245+676 \citep{Polatidis2009} and J1511+0518 \citep{An2012}, whereas others show more disturbed structures, such as 4C~31.04 \citep{Giroletti2003} and TXS~0128+554 \citep{Lister2020}.

An illustrative comparison is the radio galaxy 4C~31.04, studied by \citet{Giroletti2003}. 
This source shows a disturbed parsec-scale radio morphology, with bends, knots, and compact hotspots. 
Similar features appear in our clumpy simulations, especially in \fHigh, where jet--cloud interactions produce a fragmented distribution of bright synchrotron-emitting regions near the jet head in Figures~\ref{fig:EM-Inu} and~\ref{fig:EM-map}. 
In the simulations, these structures arise from jet deflection, shock compression, and magnetic-field amplification in the dense ambient medium. 
Depending on observing frequency, sensitivity, and angular resolution, such bright terminal structures could appear as compact hotspots or lobe-like features in young radio sources.

These comparisons suggest that the morphological differences produced by ISM clumpiness in our simulations are relevant to compact radio sources on parsec to sub-kiloparsec scales. 
Our models are idealised and are not intended to reproduce any individual object, but they show that interactions with a dense, inhomogeneous medium can generate knotty emission, compact lobe structures, and asymmetric morphologies. 
The mock maps therefore provide a bridge between the jet--ISM dynamics discussed above and observable synchrotron signatures, highlighting the role of the ambient medium in shaping the appearance of young or short-lived radio jets.

\subsection{Degeneracy with jet precession}\label{subsec:precession-clumpy}

The irregular synchrotron morphologies produced in our clumpy simulations are not unique signatures of jet propagation through a multiphase ISM. 
A similar observational appearance can also arise from intrinsic changes in the jet direction. 
For example, \textit{D24} showed that precessing relativistic jets can produce knotty and asymmetric emission structures, with bright regions displaced from the initial jet axis. 
These features are qualitatively similar to those seen in our \fHigh\ simulation, where the jet direction is fixed at injection, but interactions with dense clouds deflect the flow and produce a fragmented distribution of bright synchrotron-emitting regions.

This similarity highlights a degeneracy in the interpretation of compact radio morphologies. 
In a precessing jet, bends and off-axis knots trace changes in the injection direction, whereas in our simulations they arise from environmental deflection by an inhomogeneous ambient medium. 
In both cases, the projected emission can appear as a sequence of bright knots, compact hotspots, or lobe-like terminal structures, particularly when the source is marginally resolved or observed at a single epoch. 
Morphology alone may therefore be insufficient to determine whether a disturbed compact jet is intrinsically precessing or instead interacting with a clumpy ISM.

This ambiguity is not limited to morphology. 
For the same viewing angle, the precessing model of \textit{D24} also differs spectrally from the non-precessing homogeneous case (see fig.~9 of \textit{D24}). 
At $i=90^\circ$, the precessing jet shows a radio synchrotron peak lower than in the non-precessing homogeneous model, while developing enhanced infrared/optical emission. 
This infrared/optical increase, compared to the homogeneous case, indicates that low-level clumpiness and jet precession can produce intermediate-frequency synchrotron excesses. 
However, the physical origin is different: in the precessing case, the excess arises from the time-dependent injection direction and the redistribution of shocks and emitting particles, whereas in our simulations it is produced by jet deflection, shear, and localised shocks induced by the inhomogeneous ambient medium.

Breaking this degeneracy requires additional information. 
Multi-epoch observations can test whether bright knots follow coherent trajectories expected from a changing jet direction \citep{LuqueEscamilla2015,vonFellenberg2023}, or whether they brighten and fade locally as shocks interact with dense clouds. 
Spectral-index maps, polarisation structure, Faraday rotation, and observations of the surrounding ionised, atomic, or molecular gas can also help distinguish between intrinsic jet precession and environmental deflection. 
Our results therefore suggest that compact, asymmetric synchrotron morphologies should be interpreted with caution: they may encode either the time-dependent launching direction of the jet or the structure of the medium through which the jet propagates, and in realistic systems both effects may operate simultaneously. 
\section{Tracer Proxies for the Lepton--Baryon Content}\label{sec:leptons-baryons}

Our tracer analysis also provides a first-order proxy for how jet--ISM interaction may affect the composition of the outflow. 
In the present simulations, the passive scalars distinguish material origin rather than particle species. 
We therefore do not directly measure the lepton or baryon content of the plasma. 
However, if the injected jet is initially lepton dominated and the cloud and background gas are baryon rich, the jet tracer can be interpreted as a proxy for initially lepton-dominated material, while the ISM-origin tracer provides a proxy for baryon-rich material incorporated from the ambient medium. 
Under this assumption, the relative contribution of these two tracer components as a function of Lorentz factor gives an estimate of the baryon loading of the jet-driven outflow.

We quantify this using the cumulative tracer-weighted masses
\begin{equation}
\begin{array}{l}
\Psi_{\rm j}(\geq\gamma_f)
=
\Delta V  \, \sum_i^{\gamma_i \geq \gamma_f}
\gamma_i \rho_i \mathcal{T}_{{\rm j},i} 
, \\[1em]

\Psi_{\rm ISM}(\geq\gamma_f)
=
\Delta V  \,  \sum_i^{\gamma_i \geq \gamma_f}
\gamma_i \rho_i \left(1 - \mathcal{T}_{{\rm j},i}\right) 
\end{array}
\label{eq:tracer-content-proxies}
\end{equation}
Here, $\Psi_{\rm j}$ and $\Psi_{\rm ISM}$ are associated with jet-origin and ISM-origin material, respectively. 
The superscript on the summation indicates that the sum is restricted to cells satisfying $\gamma_i \geq \gamma_f$. 
Since the simulations use a uniform Cartesian grid, the constant cell volume $\Delta V$ is outside the summation.

\begin{figure}
\includegraphics[width=\linewidth]{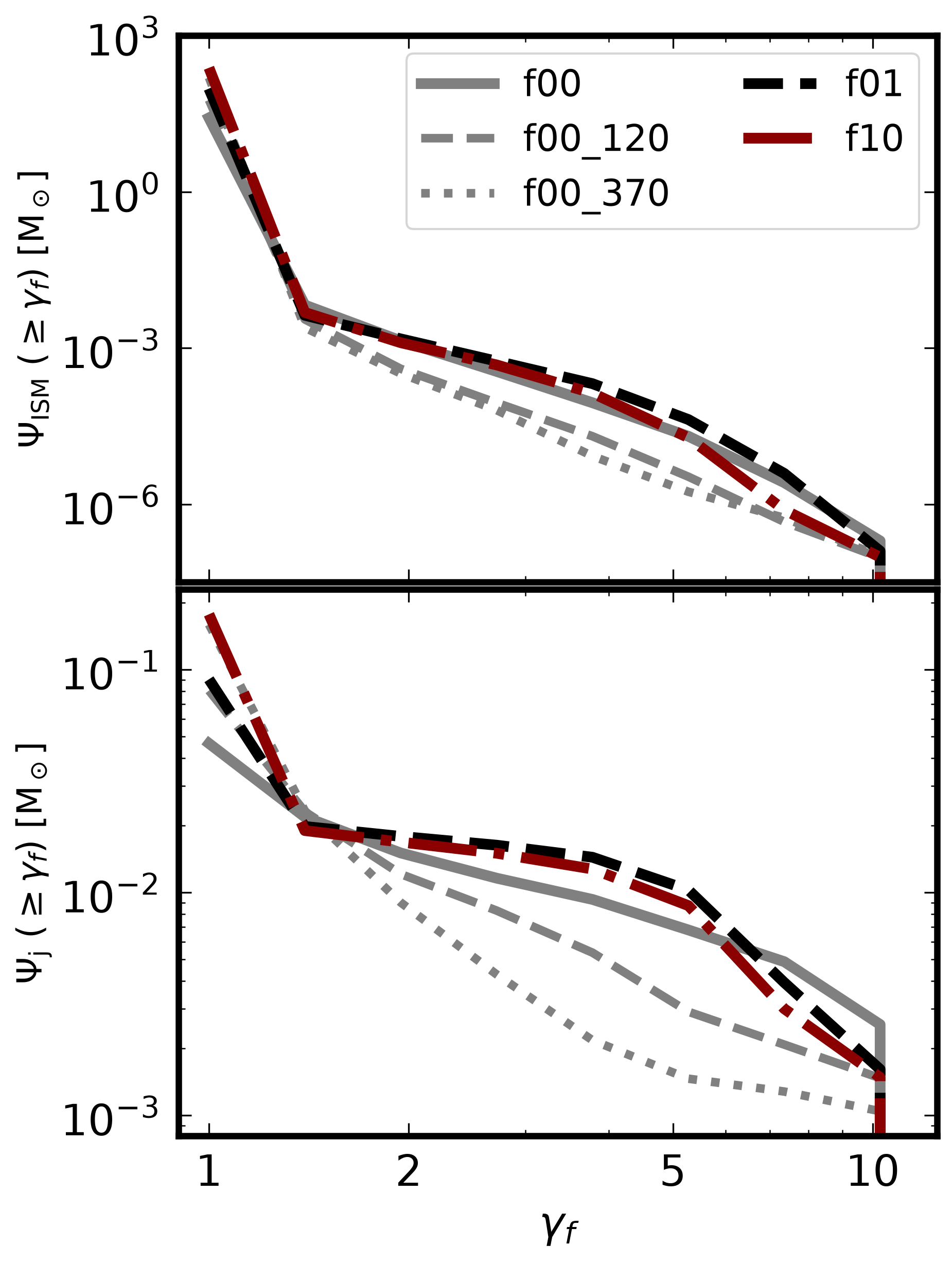}
\caption{
Cumulative tracer-weighted masses associated with ISM-origin material (top) and jet-origin material (bottom) as a function of the Lorentz-factor threshold, measured when the jets have reached comparable lengths, $z_{\rm j}\approx11$--$12$~pc.
The ISM-origin component is interpreted as a proxy for baryon-rich ISM gas, whereas the jet-origin component traces material initially injected with the jet, interpreted here as a lepton-dominated proxy.
Different colours and line styles refer to different simulations, as indicated in the legend.
The clumpy ISM modifies the cumulative distributions of both components, particularly at mildly relativistic velocities.
}
\label{fig:leptons-baryons}
\end{figure}

Figure~\ref{fig:leptons-baryons} shows the cumulative tracer-weighted masses, $\Psi_{\rm ISM}$ and $\Psi_{\rm j}$, when the jets have reached comparable lengths, $z_{\rm j}\approx11$--$12$~pc. For material with $\gamma_f\geq2$, the jet-origin contribution is substantially larger than the ISM-origin contribution in all simulations. In contrast, at $\gamma_f\simeq1$, the ISM-origin component dominates because the cumulative distribution includes the large reservoir of slow and nearly stationary ambient gas.

The clumpy ISM simulations modify both tracer distributions. Above $\gamma_f\simeq1.5$, \fLow\ and \fHigh\ show similar cumulative masses and slopes, although \fLow\ is generally slightly higher over most of the plotted range. 
The homogeneous model, \fZero, has a comparable ISM-origin distribution. 
For the jet-origin component, the clumpy simulations contain more material over $2\lesssim\gamma_f\lesssim6$, whereas \fZero\ dominates at $\gamma_f\gtrsim6$. 
This behaviour suggests that jet--cloud interactions redistribute jet-origin material from the highest Lorentz factors towards mildly relativistic velocities. Comparing simulations with the same initial total gas mass, namely \fLow\ with \fZeroB\ and \fHigh\ with \fZeroC, the clumpy cases generally contain more material above a given Lorentz-factor threshold and exhibit shallower cumulative distributions than their homogeneous counterparts, particularly for the jet-origin component over $2\lesssim\gamma_f\lesssim6$.

This difference arises from the spatial distribution of the ambient gas. In the homogeneous equal-mass simulations, the increased density provides a continuous resistance to the jet flow, promoting more coherent deceleration. 
In the clumpy simulations, low-density channels allow part of the jet-origin material to propagate while retaining mildly relativistic velocities. 
At the same time, repeated shocks, flow deflections, and cloud ablation at jet--cloud interfaces transfer momentum to localised portions of the ISM. These processes produce shallower cumulative distributions and increase the amount of both jet- and ISM-origin material at mildly relativistic velocities relative to homogeneous media with the same total gas mass. 
The decrease in jet-origin material at $\gamma_f\gtrsim6$ relative to \fZero\ further indicates that jet--cloud interactions redistribute part of the fastest jet material towards intermediate Lorentz factors.

The slightly larger cumulative masses in \fLow\ than in \fHigh\ may partly reflect differences in the porosity of the two media. Although both simulations adopt the same cloud-size distribution, the larger filling factor in \fHigh\ increases the number and projected covering fraction of clouds, reduces the typical extent of unobstructed low-density channels, and increases the total inertia of the ambient gas. The jet energy and momentum are consequently distributed over a larger gas reservoir, leaving a greater fraction of the material at low Lorentz factors. By contrast, the more porous structure of \fLow\ may allow a slightly larger amount of jet- and ISM-origin material to reach mildly relativistic velocities. However, the difference between \fLow\ and \fHigh\ is small. 
Residual differences may also arise because the selected jet lengths span $z_{\rm j}\approx11$--$12$~pc, corresponding to a mismatch of up to approximately $8$ per cent in jet length, which may affect these cumulative quantities, as well as from the particular random cloud realisations. We therefore do not interpret the small offset as evidence for a strong dependence on filling factor.

Interpreted under the composition assumption above, these trends suggest that the relativistic component remains dominated by jet-origin, lepton-dominated material, whereas the low-$\gamma_f$ budget is controlled by ISM-origin gas. 
Compared with their equal-mass homogeneous counterparts, the clumpy simulations contain more ISM-origin material at mildly relativistic velocities, indicating that jet--cloud interactions modify the incorporation and acceleration of baryon-rich material within the outflow.
However, the similarity between \fLow\ and \fHigh\ indicates that this effect does not increase strongly with cloud filling factor over the range considered. Together, the behaviour of $\Psi_{\rm j}$ and $\Psi_{\rm ISM}$ suggests that a clumpy medium can modify the partition between lepton-dominated and baryon-rich components.

We emphasise that this interpretation remains model dependent.
The simulations do not explicitly evolve separate electron and proton populations, and the non-thermal macro-particles used for the synchrotron calculation are not composition tracers.
Therefore, the inferred lepton--baryon balance depends on the assumed composition of both the injected jet and the ambient gas.
A more direct quantification would require either evolving distinct particle species, including their acceleration and cooling channels, or adopting a dedicated post-processing model that assigns a physical lepton--baryon content to each tracer component.

\section{Turbulence in the Cocoon}\label{sec:turbulence}

As the jet propagates through the ambient medium, it inflates a cocoon in which jet and ambient material become mixed. This mixing is enhanced in the presence of dense clouds, which perturb the cocoon flow and introduce additional inhomogeneities. 
The resulting fluctuations can contribute to the development of turbulence within the cocoon. We therefore analyse the turbulent properties of the cocoon material, where the effects of jet--ISM coupling and mixing are most directly captured.

We define the cocoon material using the entrained-gas selection introduced in Section~\ref{subsec:massloading}. 
To examine the density structure of this material, we calculate its density probability distribution function (PDF). 
Density fluctuations in compressible turbulent media commonly produce approximately lognormal PDFs \citep[e.g.][]{Federrath2010,Federrath2013}; therefore, the cocoon density PDF provides a useful first diagnostic of turbulence associated with the jet--ISM interaction. 

\begin{figure}
\includegraphics[width=\linewidth]{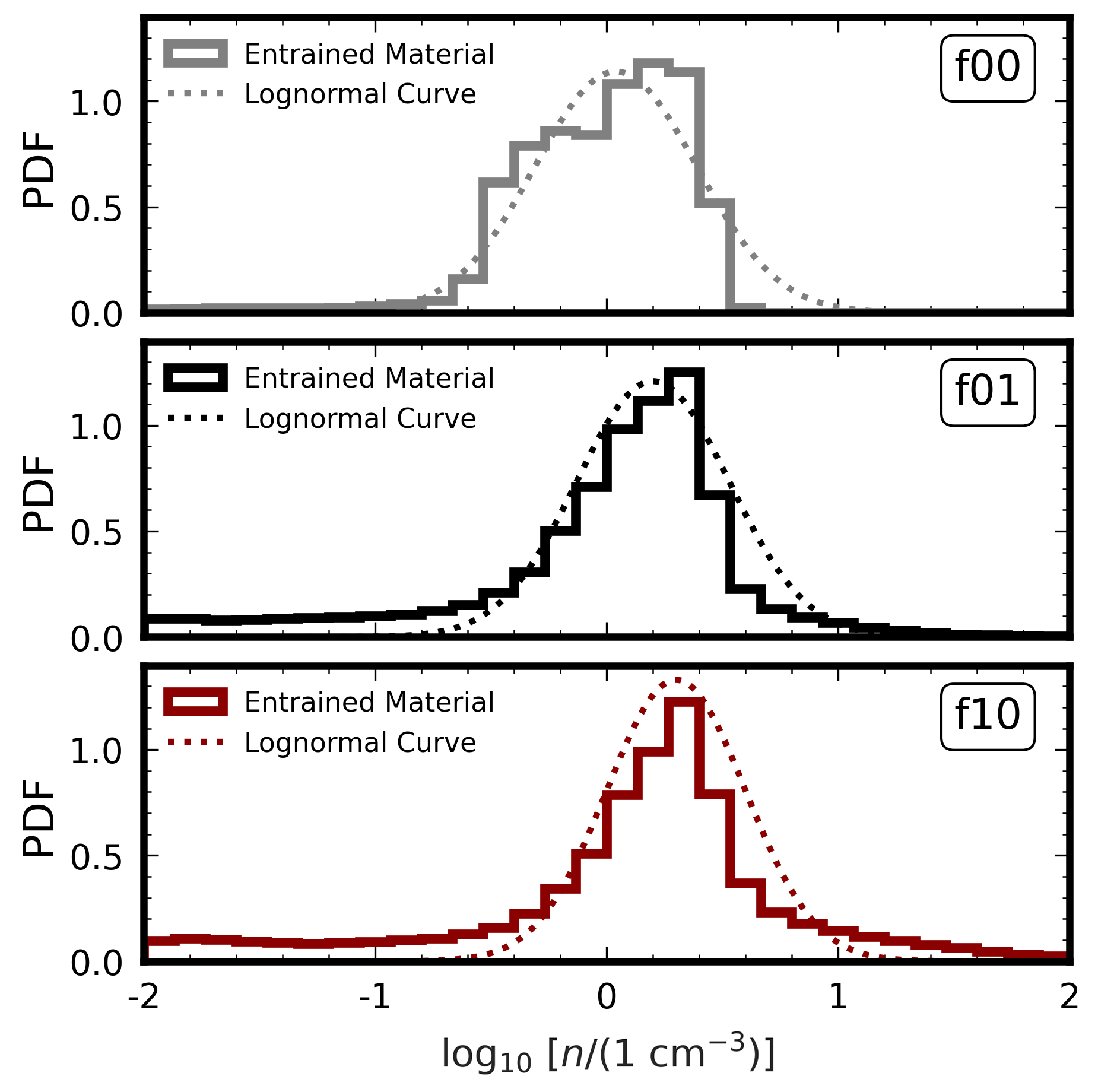}
\caption{
Density probability distribution functions (PDFs) of the cocoon material at $t=85$~yr for the three main simulations, \fZero\ (top), \fLow\ (middle), and \fHigh\ (bottom).
The solid step histograms show the simulated distributions, while the dotted curves show reference lognormal distributions included only to guide the eye, rather than formal fits.
The central parts of the distributions in the simulations with an inhomogeneous medium are broadly consistent with an approximately lognormal profile, whereas the homogeneous simulation exhibits a more structured, non-lognormal distribution.
}
\label{fig:cocoon-density-pdf}
\end{figure}

Figure~\ref{fig:cocoon-density-pdf} shows the density PDFs of the cocoon material at $t=85$~yr. In the simulations with a clumpy ISM, \fLow\ and \fHigh, the central parts of the distributions are broadly consistent with an approximately lognormal profile.
By contrast, the homogeneous simulation, \fZero, exhibits a more structured distribution, with a double-peaked central region that is not described by a single lognormal profile. The dotted curves are included only as reference lognormal distributions to guide the eye, rather than as formal fits. 
In \fLow\ and \fHigh, deviations from the approximately lognormal central distribution reflect the multiphase structure that persists within the cocoon. The low-density wings correspond to low-density channels opened by the jet during its propagation through the clumpy ISM, whereas the high-density wings are associated with the central regions of disrupted clouds that have not yet fully mixed with the surrounding cocoon gas.
Importantly, this cocoon density distribution is distinct from the initial density distribution, shown in Appenndix~\ref{app:clouds} (see Figure~\ref{fig:app-cloud-histogram}), which is dominated by the diffuse ambient component and the separate high-density cloud component. The approximately lognormal central profile therefore emerges during the jet--ISM interaction and is consistent with enhanced turbulent stirring in the clumpy cocoon.

Velocity PDFs can also be used to characterise turbulent flows \citep[e.g.][]{Banerjee2007}. 
We verified that the velocity PDFs of the cocoon material show a similar approximately lognormal behaviour and therefore provide a consistent interpretation. 
Since they do not add an independent trend to the density-PDF analysis, we do not show them separately. Instead, we complement the density PDF with locally analysed velocity and magnetic-field fluctuations, following the procedure of \textit{D23}.

To estimate the level of turbulent fluctuations in the cocoon, we follow \textit{D23} and decompose each component of the velocity and magnetic fields into a locally averaged and a fluctuating contribution. For a vector field $\boldsymbol{\mathcal{X}}$, the total local fluctuation amplitude in cell $i$ is defined as
\begin{equation}
\delta \mathcal{X}_i^T =
\left[
\sum_{k=x,y,z}
\left(
\mathcal{X}_{k,i}
-
\overline{\mathcal{X}}_{k,i}
\right)^2
\right]^{1/2},
\label{eq:turbulence-var}
\end{equation}
where $\overline{\mathcal{X}}_{k,i}$ is the mean value of component $k$ computed within a cubic region of side length $0.8$~pc, corresponding to $20$ grid cells at the adopted resolution, centred on cell $i$. We apply this decomposition to the velocity and magnetic fields of the cocoon material defined in Section~\ref{subsec:massloading}. All quantities are evaluated at $t=85$~yr, when the cocoon flow is well established in each simulation.

We quantify the relative fluctuation levels of the velocity and magnetic fields using $v_{\rm turb}/v_{\rm tot}=\delta v^{T}/|\mathbf{v}|$ and $B_{\rm turb}/B_{\rm tot}=\delta B^{T}/|\mathbf{B}|$, respectively, where $\delta v^{T}$ and $\delta B^{T}$ are obtained from the component-wise fluctuations of the corresponding vector fields. 
The median velocity fluctuation ratio increases from $0.18$ in \fZero\ to $0.29$ and $0.31$ in \fLow\ and \fHigh, respectively. 
A similar enhancement is found for the magnetic field, for which the median value of $B_{\rm turb}/B_{\rm tot}$ increases from $0.18$ in \fZero\ to $0.28$ in \fLow\ and $0.30$ in \fHigh. 
The substantially larger fluctuations in both clumpy ISM simulations indicate that the presence of dense ambient structures enhances the turbulent motions and magnetic-field disorder within the cocoon. However, the similar values obtained for \fLow\ and \fHigh\ do not indicate a clear dependence on the adopted cloud filling factor. Instead, the level of turbulence may depend non-trivially on the detailed geometry of the clumpy medium and on the resulting sequence of jet--cloud interactions.

To assess the dynamical importance of these turbulent motions, we estimate their isotropic kinetic-pressure contribution as
\begin{equation}
P_{\rm turb}
=
\frac{1}{3}\rho h
\left(\delta \mathbf{u}^{T}\right)^2.
\label{eq:P-turb}
\end{equation}
Here, $\mathbf{u} = \gamma_f\mathbf{v}$, is the spatial component of the four-velocity, and $\delta \mathbf{u}^{T}$ is calculated from equation~\eqref{eq:turbulence-var}. 
We then compute $P_{\rm turb}$ cell by cell and compare its volume-averaged value with the volume-averaged thermal pressure, $\langle P_{\rm th}\rangle$, over the cocoon.
For our adopted cocoon selection and smoothing scale, we find $\langle P_{\rm turb}\rangle/\langle P_{\rm th}\rangle \simeq 0.025$, $0.039$, and $0.061$ for \fZero, \fLow, and \fHigh, respectively.

Although this ratio increases in the clumpy ISM simulations, the turbulent-pressure contribution remains sub-dominant relative to the thermal pressure in all cases. Dense clouds therefore enhance velocity and magnetic-field fluctuations within the cocoon, but the altered jet morphology in the clumpy ISM simulations is unlikely to be driven by dominant turbulent support. Instead, it is more naturally associated with the inhomogeneous jet--cocoon interaction and the disruption of coherent large-scale flows. A more detailed characterisation of the spatial and scale-dependent turbulent energy budget is beyond the scope of the present work.

\section{Limitations and Future Steps}\label{sec:limitations}

The simulations presented here allow us to isolate the impact of a clumpy ambient medium on the early pc-scale propagation of relativistic AGN jets. 
They quantify how different cloud volume filling factors affect jet morphology, entrainment, mass loading, and synchrotron emission.
While this framework combines relativistic MHD, non-thermal particle evolution, and synchrotron-emission modelling, it relies on several simplifying assumptions that should be kept in mind when interpreting the results.

First, we do not include radiative cooling of the dense gas. 
Cooling may affect the survival of shocked clouds, the formation of dense fragments, and the thermodynamic structure of the cocoon. 
In particular, efficient cooling could allow dense gas to remain more compact after being shocked, potentially modifying the degree of mixing and the amount of entrained material. 
Second, our ISM model is intentionally idealised. 
The ambient medium is represented by a homogeneous background plus spherical dense clouds, without a self-consistent multiphase density and temperature structure, large-scale turbulence, gas self-gravity, or a cold molecular component. 
Although gravity is not expected to dominate the jet evolution over the short pc-scale timescales considered here, a background galactic potential may become relevant in longer simulations or in models that follow the ambient-gas response over larger scales. 
As a result, the simulations do not capture the full complexity of the cold and warm ISM expected in galactic nuclei.

The finite numerical resolution also limits the range of spatial scales that can be followed, even in our high-resolution simulations. 
Small-scale cloud fragmentation, turbulent mixing layers, and thin radiative structures are likely only partially resolved. 
This may affect the detailed rate of cloud disruption and mass loading, although the global trends with increasing clumpiness are likely to be less sensitive to these unresolved small-scale structures.
Higher-resolution simulations will therefore be required to test the convergence of the entrainment rates and the morphology of the shocked gas. 
In this context, adaptive mesh refinement would be particularly valuable, as it would allow dense clouds, mixing layers, shocks, and the jet--cocoon interface to be resolved more efficiently while keeping the global computational cost manageable. 
However, the particle module used in this work cannot currently be combined with the AMR implementation in \texttt{PLUTO}.

Our emission modelling is limited to synchrotron radiation. 
The electron energy distributions include IC--CMB cooling losses, but we do not compute inverse-Compton emission components. 
This includes synchrotron self-Compton emission and external Compton scattering from photon fields associated with the broad-line region, dusty torus, stellar radiation, or the extragalactic background. 
As discussed in Section~\ref{subsec:particle-injection}, external Compton scattering off the CMB is expected to be negligible for the pc-scale systems considered here, but other external photon fields may be relevant in certain configurations \citep[][]{Bhuyan2026}. 
Similarly, the lack of synchrotron self-Compton emission is not expected to affect our main conclusions, since our analysis focuses primarily on synchrotron emission from radio to X-ray frequencies rather than on the $\gamma$-ray regime. 
Nevertheless, synchrotron self-Compton emission could become relevant at higher photon energies, particularly in compact regions where the synchrotron photon energy density is high. 
The predicted spectral energy distributions should therefore be interpreted as synchrotron-dominated estimates rather than complete broadband emission models.

Future work should extend the present model towards a more realistic ISM. 
This includes incorporating radiative cooling, a broader multiphase density and temperature structure, and a cold gas component. 
Such simulations would allow a more direct assessment of how relativistic jets couple to the dense gas reservoirs observed in galactic centres.

Another natural extension is the inclusion of non-thermal protons. 
Here, we have focused on synchrotron emission from shock-accelerated electrons. 
Including proton acceleration and the associated cooling channels would make it possible to estimate the hadronic contribution to the high-energy emission. 
This would allow us to assess whether shocked jet--ISM interaction regions could contribute to high-energy non-thermal emission under plausible assumptions for proton acceleration, target-gas densities, and hadronic losses.

Overall, the present simulations provide a controlled baseline for assessing how a clumpy ISM modifies the dynamical evolution and synchrotron emission of recently launched pc-scale AGN jets. 
The extensions outlined above can build on this framework and enable more direct comparisons with compact radio sources and their high-energy counterparts.

\section{Summary and Conclusions}\label{sec:summary}

We have investigated the early dynamical and radiative evolution of pc-scale AGN jets propagating through a structured ambient media. 
We used 3D special-relativistic MHD simulations performed with the \textsc{PLUTO} code, following the first $\simeq 85$~yr of jet evolution within a $10\times10\times20$~pc domain. 
In our model, a relativistic jet with an initial Lorentz factor $\Gamma_{\rm j}=10$ is injected into an ambient medium composed of a homogeneous background gas and, in the clumpy models, a population of high-density clouds. 
We considered three ambient-medium configurations: a homogeneous medium, and two clumpy media with cloud volume filling factors of $0.1$~per cent and $1$~per cent. 
This setup allows us to isolate the role of ISM inhomogeneities in shaping the early evolution of young relativistic AGN jets.

In addition to the MHD evolution, we followed the acceleration and cooling of non-thermal electrons using Lagrangian macro-particles as in \cite{Vaidya2018}. 
This enabled us to compute synthetic synchrotron emissivity maps, mock surface-brightness maps, and integrated spectral energy distributions, linking the dynamical impact of jet--cloud interactions to their expected synchrotron signatures.

Our main results can be summarised as follows:

\begin{itemize}[leftmargin=*]

    \item[i)] \textit{Jet propagation and morphology.} 
    We find that the volume filling factor of dense clouds substantially alters the jet dynamics, even over short dynamical timescales. 
    For the same travel time, the jet head reaches shorter and shorter distances for increasing clumpiness of the ISM it penetrates.    
    This corresponds to a reduction in the mean jet-head propagation speed from $\langle v_{\rm h}\rangle\simeq0.6\,c$ for a homogeneous ISM 
    to $\simeq0.4$--$0.5\,c$ for the clumpy ISM, respectively. 
    Dense clouds slow down and deflect the jet, perturbing the flow and producing a more asymmetric morphology than in the homogeneous case.

    \item[ii)] \textit{Mass loading and entrainment.} 
    Jets propagating through a clumpy ISM experience substantially stronger mass loading and entrainment. 
    Dense clouds are shocked, stripped, and disrupted by their interaction with the jet-driven outflow, losing their initially compact morphology and contributing to the irregular structure of the cocoon. 
    This increases the amount of ambient material affected by the jet. 
    Equivalently, the entrainment-to-jet mass ratio increases from $2$ to $22$ and $48$, respectively. 
    This demonstrates that the geometrical structure of the ambient medium plays a key role in regulating the amount of ISM material swept up and entrained by the jet.
    
    \item[iii)] \textit{Material mixing and momentum transfer.}
    Applying passive tracers for the injected jet material, and the cloud material, respectively, we demonstrate that both components coexist within parts of the cocoon, 
    indicating partial mixing between jet and ISM material. 
    We have quantified this using a mass-weighted mixing factor, which increases from $2.3$~per cent for the homogeneous ISM to $4.2$~per cent and $8.9$~per cent for increasing clumpiness,
    respectively. 
    Our momentum analysis leads to the same picture: 
    most of the axial momentum remains associated with injected jet material, while only a small fraction is transferred to cloud-origin gas. 
    Thus, although clouds are disrupted and incorporated into the cocoon, this does not imply efficient acceleration or complete material homogenisation. 
    Within the simulated time-scale, cloud disruption, entrainment, acceleration, and jet--ISM mixing proceed at different rates.

    \item[iv)] \textit{Synchrotron emission.} 
    We find that the non-thermal emission is strongly affected by the presence of a clumpy ISM. 
    In general, jet--cloud interactions produce brighter, more irregular, and more spatially structured synchrotron-emitting regions than for the homogeneous case. 
    The ratio between the frequency-integrated synchrotron luminosity and the injected jet power increases from $0.15$~per cent for a homogeneous ISM  to $0.18$~per cent and
    $3.0$~per cent for increasing clumpiness, respectively, for the same injected jet (energy and mass flux). 
    The largest differences occur between the sub-mm and infrared/optical bands. 
    At $390$~THz ($\lambda\simeq770$~nm), $\nu L_{\nu}$ increases from $5 \times 10^{38}$~erg s$^{-1}$ to $10^{39}$--$\times10^{41}$~erg s$^{-1}$ for the clumpy environment, depending on the cloud filling factor. 
    We therefore propose that small-scale ISM structure can leave observable imprints on the radio-to-X-ray morphology and spectra of young jets, 
    although detailed variability diagnostics are beyond the scope of this work.

    \item[v)] \textit{Implications for compact symmetric objects.}
    The clumpy simulations produce compact, asymmetric, and knotty synchrotron morphologies that resemble some of the structural features observed in compact symmetric objects (CSOs). 
    In particular, jet--cloud interactions can generate irregular brightness distributions and apparent bends without requiring intrinsic changes in the jet-launching direction. 
    This suggests that part of the morphological diversity observed in young radio sources may arise from propagation through an inhomogeneous pc-scale ISM. 
    However, these signatures are not unique to clumpy environments, since precessing jets can produce similar radio morphologies and integrated spectra. 
    Distinguishing between these scenarios therefore requires spatially resolved, multi-frequency diagnostics rather than morphology alone.

    \item[vi)] \textit{Tracer proxies for lepton--baryon content.}
    We used the passive tracers as first-order proxies for the origin of relativistic and non-relativistic material, assuming that the injected jet is initially lepton dominated and the ambient ISM is baryon rich. 
    The cumulative tracer-weighted distributions show that high-$\gamma_f$ material is dominated by jet-origin plasma, whereas the low-$\gamma_f$ mass budget is dominated by ISM-origin material. 
    Compared with homogeneous simulations containing the same total gas mass, the clumpy models contain more jet- and ISM-origin material at mildly relativistic velocities.
    However, the similarity between the two clumpy models indicates that this effect does not increase strongly with cloud filling factor over the range considered.
    However, the tracer proxies do not by themselves provide a direct particle-by-particle composition measurement.

    \item[vii)] \textit{Turbulence in the cocoon.} 
    Jet--cloud interactions increase the level of turbulence in the cocoon and enhance the amount of shocked gas. 
    The cocoon density PDFs support this interpretation: the clumpy runs show broader, approximately lognormal distributions, with enhanced low- and high-density wings associated with jet-carved channels and disrupted cloud material. 
    Compared to the homogeneous run, the clumpy simulations develop a more irregular cocoon structure, driven by repeated jet deflections, cloud ablation, and shear at jet--cloud interfaces. 
    The median velocity-fluctuation ratio increases from $v_{\rm turb}/v_{\rm tot}=0.18$ for the case of a homogeneous ISM to $0.29$ and $0.31$ for a clumpy ISM, respectively, while the corresponding magnetic-field fluctuation ratio increases from $B_{\rm turb}/B_{\rm tot}=0.18$ to $0.28$ and $0.30$. 
    The turbulent-pressure contribution also increases with clumpiness, with $\langle P_{\rm turb}\rangle/\langle P_{\rm th}\rangle\simeq0.025$, $0.039$, and $0.061$, respectively. 
    However, we find that the turbulent pressure remains a small fraction of the total pressure, indicating that turbulence is enhanced by the inhomogeneous medium but 
    remains dynamically sub-dominant over the simulated timescale.

\end{itemize}

In summary, neglecting the inhomogeneous structure of the ambient medium can miss important aspects of the early dynamical and radiative evolution of pc-scale AGN jets. 
Even modest cloud volume filling factors modify jet propagation, morphology, entrainment, mixing, turbulence, and synchrotron emission. 
Future work including radiative cooling, a more realistic multiphase ISM, synchrotron self-Compton emission, proton acceleration, and hadronic emission will extend this framework towards more complete models of jets propagating through dense galactic environments. 

\section*{Acknowledgements}
This project was financed through a grant by the Deutsche Forschungsgemeinschaft (DFG, German Research Foundation), project number 443220636, via the Research Unit FOR\,5195 (see \href{www.for5195.uni-wuerzburg}{www.for5195.uni-wuerzburg}).
BV would like to acknowledge the financial support of the German Academic Exchange Service (DAAD) under the Research Grants -- Bi--nationality Supervised Doctoral Degrees/Cotutelle, 2025/26 (57730834), which facilitated the travel and research collaboration between our institutions.

We acknowledge enlightening conversations with Harshita Bhuyan, Aditya Sharma, Anindya Guria, Tom Weidemeier, and Karl Mannheim.
IA acknowledges useful discussions with Rogemar André Riffel, Raniere de Menezes, Rodrigo Nemmen, Chris Harrison, and Stephane V. Werner.

We thank the anonymous referee for their constructive comments, which helped improve the clarity of the manuscript.

The \texttt{PYTHON} package \citep{Millman2011} has been used for data analysis and figure production.
This research also benefited from the \texttt{PyPLUTO} package for data visualisation \citep{PyPLUTO2025}. We further used \texttt{numpy} \citep{Numpy}, \texttt{matplotlib} \citep{Matplotlib}, \texttt{pandas} \citep{reback2020, mckinney-proc-scipy-2010}, and \texttt{scipy} \citep{Virtanen2020}.

Computations were performed on the HPC systems Raven and Viper at the Max Planck Computing and Data Facility.

\section*{Data Availability}

The data underlying this article will be shared on reasonable request to the corresponding author.



\bibliographystyle{mnras}
\bibliography{refs} 




\appendix

\section{Jet injection profile}\label{app:jet-profile}
In the nozzle region, we initialise the jet using the relativistic rotating magnetised equilibrium solution derived by \citet{Bodo2019}. The adopted profiles are
\begin{equation}
\begin{array}{l}
n(r) =
n_{\rm bg}
-
\displaystyle\frac{n_{\rm bg}-n_{\rm j}}
{\cosh\left[\left(r/r_j\right)^8\right]},
\\[1em]

\gamma_z(r) =
1+
\displaystyle\frac{\gamma_c-1}
{\cosh\left[\left(r/r_j\right)^6\right]},
\qquad
v_z(r)=\sqrt{1-\gamma_z^{-2}},
\\[1em]

v_\phi^2(r) =
\frac{r^2 \gamma_c^2 \Omega_c^2}{\gamma_z^2}
\left[
1+r^2\gamma_c^2\Omega_c^2
\exp\left(-\frac{r^4}{a^4}\right)
\right]^{-1}
\exp\left(-\frac{r^4}{a^4}\right),
\\[1em]

v_r = B_r = 0,
\\[0.5em]

a^{4} B_{zc}^{2} =
\frac{H_{c}^{2} \delta_{c}^{2}}
{1-\left(\delta_{c}\Omega_{c}+v_{zc}\right)^{2}},
\\[0.5em]

\zeta \equiv
\frac{\rho_{j}\gamma_{c}^{2}\Omega_{c}^{2}a^{4}}
{2H_{c}^{2}},
\\[0.5em]

B_z^2 =
B_{zc}^2 -
(1-\zeta)
\frac{H_c^2\sqrt{\pi}}{a^2}
\mathcal{E}\left(\frac{r^2}{a^2}\right),
\\[0.5em]

B_\phi(r) =
\frac{
- v_\phi v_z B_z -
\sqrt{v_\phi^2 B_z^2 + H^2(1-v_z^2)}
}
{1-v_z^2}.
\end{array}
\label{eq:Bodo-equations}
\end{equation}

 $\Omega$ is the angular velocity, $a$ is the magnetisation radius,
$H^2\equiv B_\phi^2-E_r^2$, $E_r=v_zB_\phi-v_\phi B_z$, and
$\delta=rB_z/B_\phi$ is the magnetic pitch angle. The function $\mathcal{E}$ denotes the error function, and the subscript ``$c$'' indicates central values at $r=0$.

Here, $n_{\rm j}$ is the hydrogen number density on the jet axis. The density transitions towards the background value, $n_{\rm bg}$, near the nozzle boundary. This transition accounts for the finite width of the low-density component in the initial density distribution shown in Figure~\ref{fig:app-cloud-histogram}. The quantity $\gamma_z$ is the Lorentz factor associated with the axial velocity component, $v_z$, and exhibits a smooth but steep radial decrease from $\gamma_c$ on the jet axis. The total Lorentz factor is calculated from both $v_z$ and the azimuthal velocity component, $v_\phi$.
Furthermore, $\Omega$ is the angular velocity, $a$ is the magnetisation radius,
$H^2\equiv B_\phi^2-E_r^2$, $E_r=v_zB_\phi-v_\phi B_z$, and
$\delta=rB_z/B_\phi$ is the magnetic pitch angle. The function $\mathcal{E}$ denotes the error function, and the subscript ``$c$'' indicates central values at $r=0$.

This solution enforces transverse equilibrium within the injected beam. The velocity and magnetic-field profiles are therefore specified consistently, rather than by prescribing independent rotation and toroidal-field profiles. 
In the simulations, these expressions are applied only inside the cylindrical injection nozzle, where the primitive variables are reset at each timestep to impose a steady inflow. 
We choose the ambient thermal pressure to match the jet thermal pressure, $p_{\rm bg}=p_{\rm j}$, so that the injected beam is not initially over- or under-pressurised with respect to its surroundings the same way as in \textit{D23}. 
For the full derivation of the transverse equilibrium solution, we refer to \citet{Bodo2019}.


\section{Synchrotron calculation} \label{app:synchrotron}

The energy distribution of each macroparticle evolves as it is advected through the jet and shocked ambient medium. We account for adiabatic losses associated with the flow expansion, radiative losses from synchrotron emission, and inverse Compton (IC) losses due to scattering of CMB photons. 
The IC--CMB losses are subdominant for the redshift and time-scales considered here, but are retained for consistency with the particle-evolution model. When macroparticles cross shocks, their electron distributions can also be modified by diffusive shock acceleration, following the implementation described in \textit{D23} and \textit{D24}. 
In this work, we compute only the synchrotron emissivity and do not include IC--CMB or synchrotron self-Compton
emission.

The synchrotron emissivity is calculated following the method described in \textit{D24} (see their Section~2.3), based on the formalism of \citet{Ginzburg1965}. The comoving-frame emissivity produced by relativistic electrons moving along the line of sight $\hat{n}'_{\mathrm{los}}$, between
minimum and maximum energies $E_0$ and $E_1$, is
\begin{equation}
\begin{split}
    {J}'_{\mathrm{syn}}(\nu', \hat{n}'_{\mathrm{los}}, \mathbf{B}')
    = \frac{\sqrt{3}\, e^3}{4\pi\, m_e c^2}
    \left| \mathbf{B}' \times \hat{n}'_{\mathrm{los}} \right| \\
    \times \int_{E_0}^{E_1} \mathcal{N}'(E')\, F(x)\, dE',
\end{split}
\label{eq:syn_emissivity}
\end{equation}
where primes denote quantities measured in the comoving frame, $\nu'$ is the emitted frequency, $\mathbf{B}'$ is the local magnetic field, $e$ is the electron charge, $m_e$ is the electron mass, and $E'$ is the particle energy. The function $F(x)$ is defined as
\begin{equation}
\begin{split}
    F(x) = x \int_{x}^{\infty} K_{5/3}(z)\, dz,  \quad x \equiv \frac{4\pi\, m_e^3 c^5 \nu'}{3 e {E'}^{2} \left| \mathbf{B}' \times \hat{n}'_{\mathrm{los}} \right|},
\end{split}
\label{eq:F_function}
\end{equation}
where $K_{5/3}$ is the modified Bessel function of the second kind. The emissivity in the observer frame is obtained through
\begin{equation}
{J}_{\mathrm{syn}} = \mathcal{D}^2 {J}'_{\mathrm{syn}},
\qquad
\mathcal{D} = \frac{1}{\gamma_f(1 - \vec{\beta}\,\cdot\,\hat{n}_{\mathrm{los}})},
\label{eq:doppler_emissivity}
\end{equation}
where $\mathcal{D}$ is the Doppler factor, $\vec{\beta}=\mathbf{v}/c$, and $\gamma_f$ is the fluid Lorentz factor. For further details of the particle evolution and emissivity calculation, we refer to \textit{D24} and
\citet{Vaidya2018}.

The synchrotron emissivity is computed for each macroparticle and then interpolated onto the numerical grid described in Section~\ref{subsec:code}. We define $J_\nu(x,y,z)$ as the sum of ${J}_{\mathrm{syn}}$ over all
macroparticles located inside the cell centred at $(x,y,z)$.

\section{Clouds in the Initial Condition}\label{app:clouds}

\begin{figure}
\includegraphics[width=0.95\linewidth]{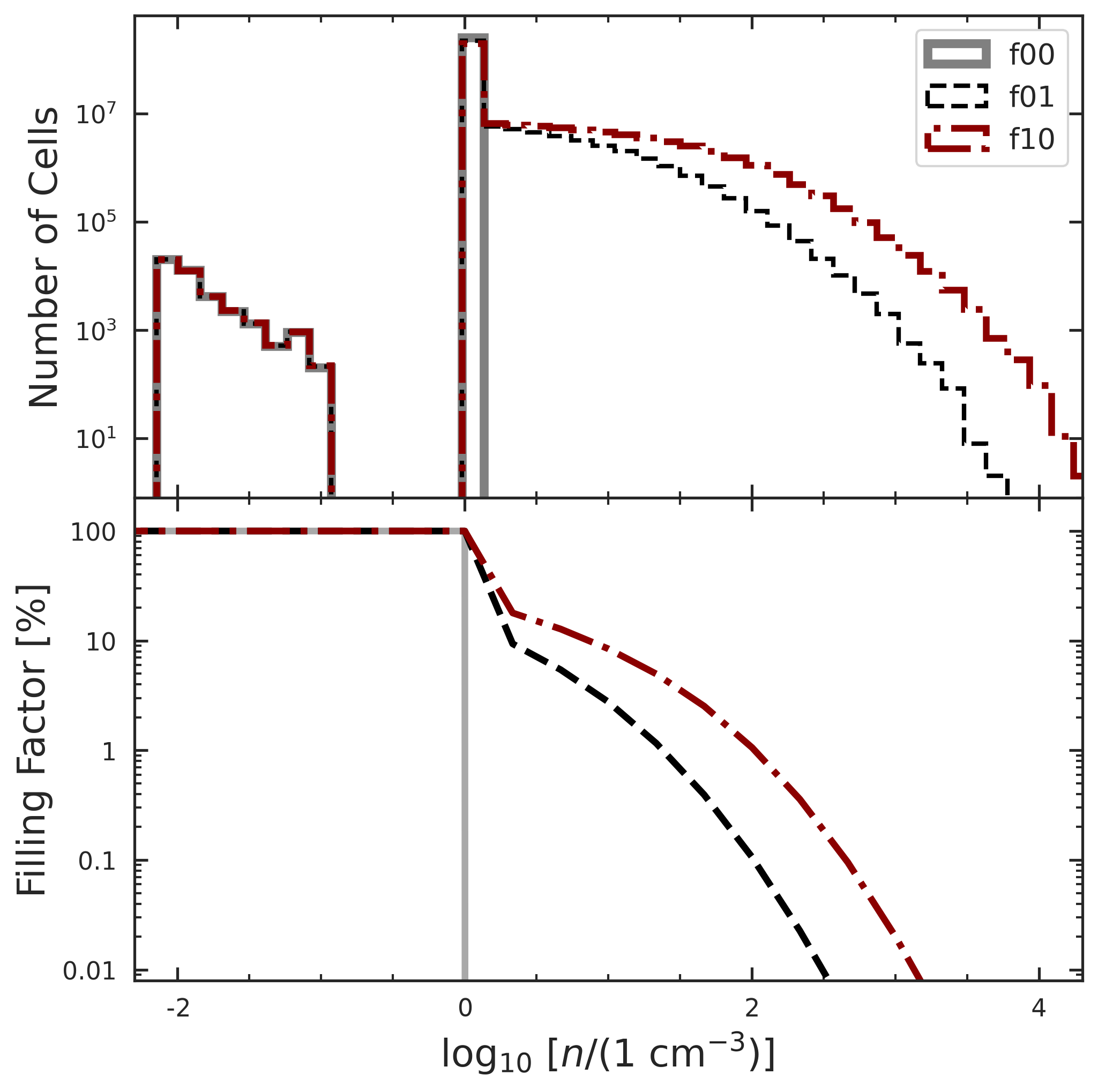}
\caption{
Top: Distribution of hydrogen number density in the initial conditions for our main simulations. The histogram shows the imposed multi-component initial density structure, with three main components: the low-density jet ($n < 1 ~\dens$), the dominant background gas, centred at $n=1\,\dens$, and the dense gas associated with the clouds ($n>10^2\,\dens$). 
Bottom: Volume filling factor, $f$, as a function of density, computed using Equation~\ref{eq:filling-factor}. The filling factor decreases with increasing density, reflecting the small volume occupied by the densest cloud cores compared to the more diffuse background medium. The quoted filling factor is evaluated at $n=10^2\,\dens$.}
\label{fig:app-cloud-histogram}
\end{figure}

\begin{figure}
\includegraphics[width=\linewidth]{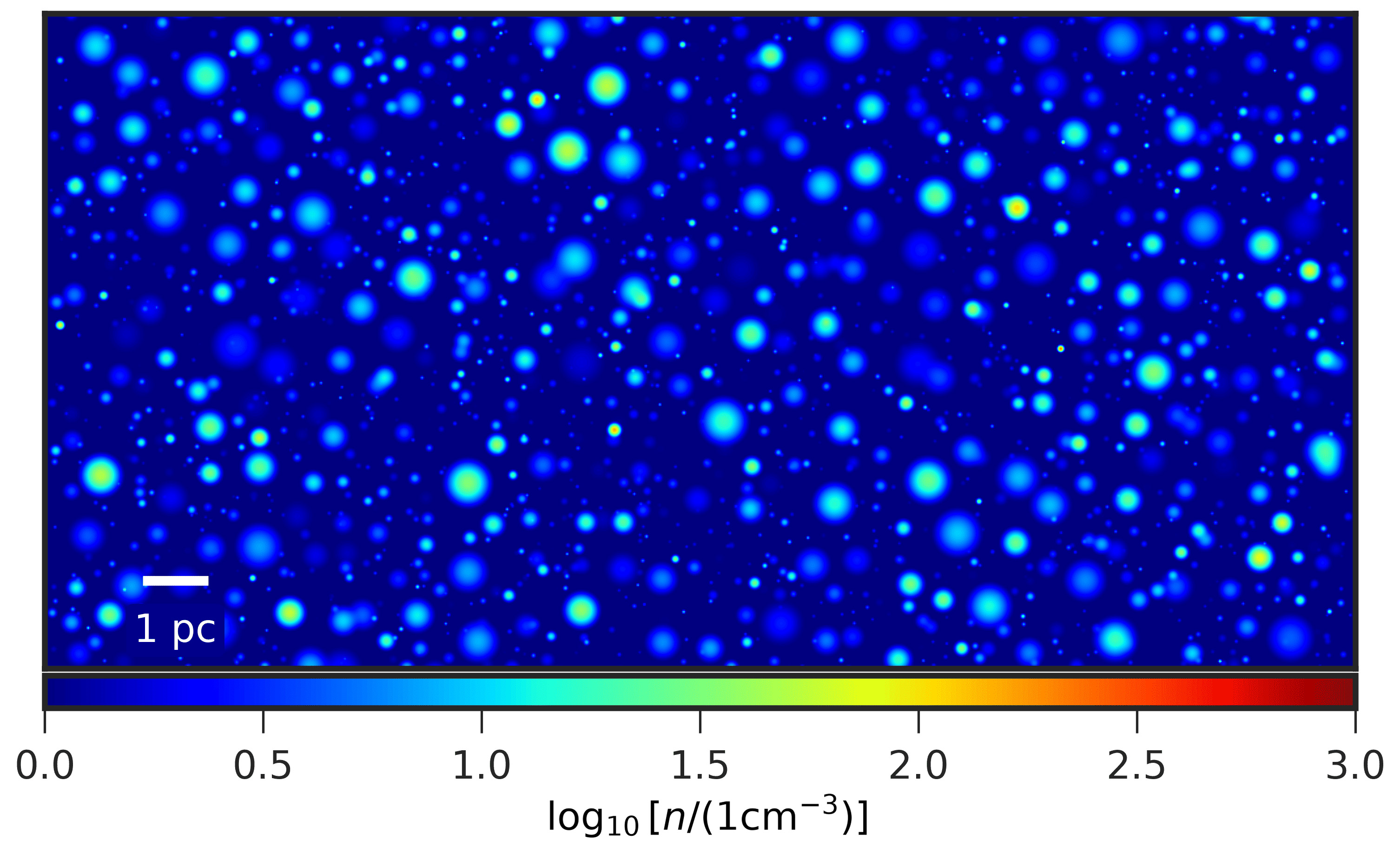}
\caption{
Two-dimensional slice of the number density for the \fLow\ simulation, shown for a projection of width $1$~pc.
The map illustrates the initial distribution of randomly placed spherical clouds embedded in the uniform background medium.
The cloud radii span $R_{\rm cl}=0.05$--$1$~pc, and their central densities are drawn from the lognormal distribution described in Appendix~\ref{app:clouds}.
}
\label{fig:app-cloud-density}
\end{figure}

The cloud centres are randomly distributed within the computational domain, excluding the jet-injection nozzle. 
Each cloud has radius $R_{\rm cl}$ and a central density $n_c$. 
The cloud radii are sampled in the range $R_{\rm cl}=0.05$--$1$~pc from a power-law distribution, $N(R_{\rm cl})\propto R_{\rm cl}^{-2}$. 
The central density $n_c$ is drawn from a lognormal distribution centred at $n_0=10^{3.5}\,\dens$, with standard
deviation $\sigma=0.5$, and restricted to the range $n_c=10^2$--$10^5\,\dens$. This range is consistent with dense-gas estimates for nearby active galaxies \citep{Davies2020, Revalski2025}.

Within each cloud, the density decreases smoothly from $n_c$ at the centre to the background density $n_{\rm bg}$ at the cloud boundary. The radial profile is given by
\begin{equation}
n_{\rm cl}(r') = n_c \, 10^{- \frac{r'^2}{R_{\rm cl}^2}\log_{10} \left( \frac{n_c}{n_{\rm bg}} \right)}.
\label{eq:n_cloud}
\end{equation}
where $r'$ is the distance from the selected cloud centre. 
This prescription ensures that $n_{\rm cl}(0)=n_c$ and $n_{\rm cloud}(R_{\rm cl})=n_{\rm bg}$. Thus, the
clouds are smoothly matched to the diffuse background and are never underdense relative to the ambient medium.

The clouds are initially at rest and in thermal pressure equilibrium with the diffuse ambient gas, such that $p_{\rm cl}=p_{\rm bg}$, to avoid artificial early cloud expansion or collapse.
This choice causes the cloud temperature to vary inversely with density, $T_{\rm cl}(r')\propto p_{\rm cl}/n_{\rm cl}(r')$.
The cloud centres are therefore cooler than their outskirts. However, because the density profile satisfies $n_{\rm cl}(R_{\rm cl})=n_{\rm bg}$, the temperature also approaches the diffuse-background value at the cloud boundary.
Thus, the cloud--background interface does not contain a temperature jump, although the radial temperature gradient changes where the cloud joins the uniform background.
We verified the stability of the initial cloud distribution with a control simulation evolved without jet injection for $\sim100$~yr, comparable to the duration of the main simulations. The cloud structure remains stable over this time-scale.
The assumption of initially stationary clouds is adequate for the present numerical experiment. 
Even velocities comparable to a high galactic velocity dispersion, $v_{\rm cloud}\lesssim1000$~km~s$^{-1}$, would move the clouds by at most $\sim0.1$~pc over $\sim100$~yr, which is negligible compared with the scales of interest.

For each target filling factor, the cloud population is normalised such that the volume occupied by gas with $n>10^2\,\dens$ satisfies equation~\eqref{eq:filling-factor}. 
The resulting density distributions are shown in the top panel of Figure~\ref{fig:app-cloud-histogram}. The histograms show three components: the low-density jet, the diffuse background medium, and the dense gas associated with the clouds. 
The bottom panel shows the cumulative volume filling factor as a function of density threshold; the value quoted for each simulation corresponds to the threshold $n=10^2\,\dens$.

Figure~\ref{fig:app-cloud-density} shows a density slice of width $1$~pc for the \fLow\ simulation. The figure illustrates the randomly distributed spherical clouds embedded in the diffuse background medium.
The simulations use a uniform Cartesian grid with spatial resolution $\Delta x\simeq0.02$~pc. The smallest clouds are therefore resolved by only a few cells across their radius, whereas the largest clouds are resolved by several tens of cells. 
Although the smallest clouds are only marginally resolved, the aim of this work is not to study the detailed evolution of individual clouds, but to capture the global interaction between a relativistic jet and a clumpy medium. 
A detailed study of individual cloud disruption would require higher resolution in the dense gas and additional physics, such as radiative cooling and self-gravity.


\bsp	
\label{lastpage}
\end{document}